\documentclass[superscriptaddress,twocolumn,showpacs]{revtex4}
\usepackage{graphicx}
\usepackage{color}
\usepackage{dcolumn}
\usepackage{amsmath}
\usepackage{dsfont}

\newcommand{\be}{\begin{equation}}
\newcommand{\ee}{\end{equation}}

\begin{document}

\title{Superdiffusive Transport and Energy Localization in Disordered Granular Crystals}

\author{Alejandro J. Mart\'inez}

\affiliation{Oxford Centre for Industrial and Applied Mathematics,
Mathematical Institute, University of Oxford, Oxford OX2 6GG, UK}

\author{P. G. Kevrekidis}

\affiliation{Center for Nonlinear Studies and Theoretical Division, Los Alamos
National Laboratory, Los Alamos, NM 87544}

\affiliation{Department of Mathematics and Statistics, University of
Massachusetts, Amherst, Massachusetts 01003-4515, USA}

\author{Mason A. Porter}

\affiliation{Oxford Centre for Industrial and Applied Mathematics,
Mathematical Institute, University of Oxford, Oxford OX2 6GG, UK}

\affiliation{CABDyN Complexity Centre, University of Oxford, Oxford OX1 1HP, UK}




\begin{abstract}

We study the spreading of initially localized excitations in one-dimensional disordered granular crystals. We thereby investigate localization phenomena in strongly nonlinear systems, which we 
demonstrate to be fundamentally different from localization in linear and weakly nonlinear systems. We conduct a thorough comparison of wave dynamics in chains with three different 
types of disorder: an uncorrelated (Anderson-like) disorder and two types of correlated disorders (which are produced by random dimer arrangements), and for two families of initial conditions: 
displacement perturbations and velocity perturbations.
We find for strongly precompressed (i.e., weakly nonlinear) chains that the dynamics strongly depends on the initial condition. In particular, for displacement perturbations, the  
long-time asymptotic behavior of the second moment $\tilde{m}_2$ has oscillations that depend on the type of disorder, with a complex trend that is markedly different from a power law and which is particularly evident for an Anderson-like disorder. By contrast, for velocity perturbations, we find that a standard scaling $\tilde{m}_2\sim t^{\gamma}$ (for some constant $\gamma$) applies for all three types of disorder. For weakly precompressed (i.e., strongly nonlinear) chains, 
$\tilde{m}_2$ and the inverse participation ratio $P^{-1}$ satisfy scaling relations $\tilde{m}_2\sim t^{\gamma}$ and $P^{-1}\sim t^{-\eta}$, and the dynamics is superdiffusive for all of the cases that we consider. Additionally, when precompression is strong, the inverse participation ratio decreases slowly (with $\eta<0.1$) for all three types of disorder, and the dynamics leads to a partial localization around the core and the leading edge of the wave. For an Anderson-like disorder, displacement perturbations lead to localization of energy primarily in the core, and velocity perturbations cause the energy to be divided between the core and the leading edge. This localization phenomenon does not occur in the sonic-vacuum regime, which yields the surprising result that the energy is no longer contained in strongly nonlinear waves but instead is spread across many sites. 
In this regime, the exponents are very similar (roughly $\gamma\approx 1.7$ and $\eta\approx 1$) for all three types of disorder and for both types of initial conditions.

\end{abstract}

\maketitle



\section{Introduction}

The study of wave propagation in disordered lattice and continuum systems
has been an important and popular research theme during the past several decades. Some of the most prominent recent studies on these topics~\cite{Schwartz:Nature2007,Billy:Nature2008,Roati:Nature2008,Gurarle:PRL2008,Conti:NatPhys2008,Efremidis:PRL2008,Hu2008,Gremaud:PRL2010} have generalized to weakly nonlinear settings the ideas of P.~W. Anderson, who showed theoretically that the diffusion of waves is curtailed in linear random media (where the randomness arises from defects or impurities) \cite{Anderson:PR1958,Kramer:RPP1993}.  
This interplay between disorder and nonlinearity --- which often arises in the 
presence of lattice discreteness --- is of considerable interest to 
a vast array of ongoing studies, as is evidenced by the
recent reviews~\cite{Flach:ArxivRep2014-1,Flach:ArxivRep2014-2} (see 
also the numerous references therein). The set of different physical scenarios in which Anderson localization has been investigated is staggering: it ranges all the way from electromagnetism \cite{Schwartz:Nature2007} and acoustics \cite{Hu2008} to subjects such as quantum chromodynamics \cite{anderson-qcd}. 

As in the above studies, we are interested in waves in disordered media, but we depart from the earlier work in a very important way: we seek to explore order--disorder transitions with a particular emphasis on \textit{strongly nonlinear} media.  This contrasts sharply with the linear and weakly nonlinear media in which Anderson-like models have traditionally been studied~\cite{Flach:ArxivRep2014-1,Flach:ArxivRep2014-2}. Our approach is motivated predominantly by the strong (and increasing) interest in granular crystals~\cite{Nesterenko:book,Sen:PR2008,pgk:2011}, which (as we discuss below) are very important both for the study of fundamental nonlinear phenomena and for numerous engineering applications. The examination of disordered systems in general --- and of Anderson-like phenomena in particular --- is a key challenge in the study of nonlinear chains~\cite{Kuzovkov:PhysScr2011,Skokos:PRE2009,Mulansky:PRE2009}.

One-dimensional (1D) granular crystals, which consist of closely packed chains of elastically colliding particles, are a paradigmatic system for the investigation of chains of strongly
nonlinear oscillators. Their strongly nonlinear dynamic response has inspired numerous studies of
the interplay between nonlinearity and discreteness \cite{Nesterenko:book,Sen:PR2008,pgk:2011}.  One can construct granular crystals using materials of numerous types and sizes, and their properties are thus extremely tunable, tractable, and flexible \cite{Nesterenko:book,Sen:PR2008,Daraio:PRE2006,Coste:PRE1997}.  
This also makes them very well-suited for investigating the effects of structural and material heterogeneities on nonlinear wave dynamics. Studies have examined the role of defects \cite{Sen:PRE1998,PhysRevE.57.2386,Hascoet:EPJB2000,Hong:APL2002,Theocharis:PRE2009} (including in experimental 
settings~\cite{Job:PRE2009,Many}), interfaces between 
two different types of particles 
\cite{Nesterenko:PRL2005,Daraio:PRL2006}, decorated and/or tapered chains \cite{Doney:PRL2006,Harbola:PRE2009}, chains of diatomic and triatomic units \cite{Porter:PRE2008,Porter:PhysicaD2009,Herbold:Mechanica2009,Molinari:PRE2009,staros1,staros2,staros3,boechler4}, and quasiperiodic and random 
configurations \cite{Sokolow:AnnPhys2007,Chen:PhysicaB2007,Fraternali:MAMS2010}. The tunability of
granular crystals is valuable not only for fundamental studies of their underlying physics but also in potential engineering applications --- including shock and energy absorbing layers 
\cite{Daraio:PRL2006,Hong:PRL2005,Fraternali:MAMS2010}, sound focusing devices and delay lines \cite{Spadoni:PNAS2010}, actuators \cite{Khatri:SPIE2008}, 
vibration absorption layers \cite{Herbold:Mechanica2009}, sound scramblers \cite{Daraio:PRE2005,Nesterenko:PRL2005}, and acoustic switches and logic
gates~\cite{daraio_natcom}.  Because one can model granular chains as a type of Fermi-Pasta-Ulam (FPU) lattice, they have also been employed in studies of phenomena such as equipartition (see, e.g., \cite{szel2013,zhang2015}).

As was illustrated recently, localization in strongly nonlinear systems can have a fundamentally different character from localization in linear and weakly nonlinear systems \cite{Ponson:PRE2010}.  
Importantly, one can use the setting of granular crystals to explore a regime (the so-called ``sonic vacuum'') in which no linear counterpart whatsoever exists~\cite{Nesterenko:book}. It is our goal in the present 
paper to investigate this regime and nearby regimes in detail and to conduct what we believe is the first systematic study of the differences between localization is linear, weakly nonlinear, and strongly nonlinear systems. 
There are numerous types of disorder in a granular chain, and --- as we demonstrate in this paper --- 
it matters whether the disorder is uncorrelated (as in the original Anderson model) or correlated.  Moreover, there are multiple types of possible correlations in disordered arrangements, and we illustrate using randomized arrangements of dimers (see Ref.~\cite{Ponson:PRE2010} for an example arrangement that was studied in the context of granular crystals) that seemingly small differences in disorder can have a large impact
on the dynamics of wave propagation in strongly nonlinear systems. 
Moreover, because granular chains are a type of FPU system \cite{FPU} --- so nonlinearities arise from the potentials that connect adjacent nodes of the lattice --- they are fundamentally different from the nonlinear Schr{\"o}dinger (NLS) and Klein--Gordon (KG) lattices in which disordered configurations have been extensively studied 
recently~\cite{Flach:ArxivRep2014-1,Flach:ArxivRep2014-2}. 
In fact, as we will demonstrate in the present paper, this difference leads to much more rapid transport in disordered granular chains than what occurs in either NLS or KG lattices.  This fundamental difference is one of the main findings of our work: the dynamics of strongly nonlinear, disordered granular crystals includes regimes with superdiffusive transport.  

Much of the significant volume of work involving nonlinear disordered lattices has focused on the effect of weak nonlinearity on the well-established Anderson model~\cite{Flach:ArxivRep2014-1,Flach:ArxivRep2014-2,Pikovsky:PRL2008,Tietsche:EPL2008,Johansson:EPL2010,Laptyeva:EPL2010,Garcia-Mata:PRE2009}. One of the most remarkable findings in this body of work
is the fact that a small amount of nonlinearity in a disordered lattice can induce interaction between Anderson modes, which eventually can lead to a subdiffusive delocalization process. Interestingly, this phenomenon emerges in a highly nontrivial way: even when both disorder and nonlinearity separately tend to localize energy, they also ``interfere'' with each other's transport-generation processes and consequently destroy the Anderson-localization mechanism. 
In the context of strongly nonlinear disordered lattices, a noteworthy
recent effort is that of~\cite{Mulansky:NJP2013}. The authors of that paper examined disordered lattices --- where disorder is introduced via either a linear or a nonlinear on-site term --- in which the coupling leads to a strongly nonlinear setting. They found that initially localized wave packets tend to spread in a subdiffusive way. 
However, the spreading in nonlinearly coupled linear oscillators is slow in comparison to purely subdiffusive behavior. We believe that the subdiffusive behavior in their setting is a consequence of the local potential, because (as we demonstrate in our paper) the dynamics tends to be superdiffusive when only strongly nonlinear interactions are present.\footnote{We note in passing that superdiffusive behavior was reported very recently in a so-called ``pseudo-two-dimensional'' random 
dimer~\cite{rodrigo}. Both the setting and qualitative behavior of the system in Ref.~\cite{rodrigo} are different from ours in fundamental ways. They consider purely linear dynamics, whereas we consider both linear and (especially) nonlinear dynamics. Additionally, they found superdiffusive transport for lattices with short-range correlations but subdiffusive transport for uncorrelated disorders (such as the one in the Anderson model), whereas we find superdiffusive transport in lattices with either correlated or uncorrelated disorder.}

The remainder of our paper is organized as follows. In Sec.~\ref{Sec3}, we describe the fundamental equations that characterize a disordered granular chain with Hertzian interactions, and we examine 
different approximations that depend on the amount of precompression. In Sec.~\ref{types}, we present three different types of disorder and study their correlation properties, which will prove to 
be of crucial importance for the qualitatively different transport dynamics that can occur. In Sec.~\ref{Sec4}, we briefly discuss the influence of impurities in homogeneous chains on the modes that emerge and on the dynamics more generally.
In Sec.~\ref{Sec5}, we present our computational results. We describe the fundamental differences between the different types of disorder, and we discuss the properties of the linear 
spectrum and the different types of impurity-like modes that appear
for each type of disorder. We also study the transport and localization 
properties for both linear and nonlinear waves for each type of disorder, and we demonstrate with numerical simulations that localization (as either breathers or traveling waves) is no longer 
possible in a strongly nonlinear regime.
Additionally, for strongly precompressed chains and for initially localized displacement excitations, we demonstrate that the second moment exhibits a complicated trend that differs markedly from a 
power law. By contrast, we observe superdiffusive transport for all of the other configurations and initial conditions.
We summarize our conclusions and discuss future challenges in Sec.~\ref{Sec6}.


\section{Disordered Granular Crystals}\label{Sec3}

\subsection{Equations of Motion}

One can describe a 1D crystal of $N$ spherical particles as a chain of
nonlinear coupled oscillators with Hertzian interactions between
each pair of particles \cite{Nesterenko:book,Sen:PR2008,pgk:2011}.  
Hertzian forces are applicable to a wide variety of 
materials~\cite{Johnson:Book} (including steel, aluminum,
brass, bronze, and many more). The equations of motion in this setting are
\begin{equation}
	\ddot{u}_n = \frac{A_n}{m_n}[\Delta_n+u_{n-1}-u_n]_+^{3/2}-
	\frac{A_{n+1}}{m_n}[\Delta_{n+1}+u_n-u_{n+1}]_+^{3/2}\,,
\label{Hertz}
\end{equation}
where $u_n$ is the displacement of the $n$th particle (where $n\in \{1,2,{\ldots},N\}$) measured from its equilibrium position in the initially compressed chain, $m_n$ is the mass of the $n$th particle, and 
\begin{equation}
	\Delta_n = \left(\frac{F_0}{A_n}\right)^{2/3}
\end{equation}
is a static displacement for each particle that arises from the static load $F_0 =
\text{const}$.  The parameter $A_n$ is given by
\begin{equation}
	A_n =
	\frac{4E_{n-1}E_{n}\left(\frac{R_{n-1}R_n}{R_{n-1}+R_n}
	\right)^{1/2}}{3\left[E_{n}(1-\nu_{n-1}^2)+E_{n-1}(1-\nu_{n}^2)\right]}\,,\label{An}
\end{equation}
where the elastic modulus of the $n$th particle is $E_n$, the Poisson ratio of the $n$th particle is $\nu_n$, and the radius of the $n$th particle is $R_n$. A Hertzian interaction between a pair of particles occurs only when they are in contact, so each particle is affected directly only by its nearest neighbors and experiences a force from a neighbor only when it overlaps with it. This yields the bracket
\begin{equation}
[x]_+ = \left\{
	\begin{array}{lcc}
		x\,, & \text{if} & x>0\\
		0\,, & \text{if} & x\leq0
	\end{array}\right.
\end{equation}
in Eq.~\eqref{Hertz}. The exponent $3/2$ and prefactor $A_n$ in Eq.~\eqref{Hertz} are consequences of the elastic nature of the particle interactions and of the particle geometry~\cite{Nesterenko:book,Sen:PR2008}. 
Other particle shapes, such as ellipsoids~\cite{Ngo:PRE2011} and cylinders~\cite{Khatri:Granular2012}, can also exhibit Hertzian interactions.  

The boundary conditions of Eq.~\eqref{Hertz} are given by considering $u_0=u_{N+1}=0$ and $R_0,R_{N+1}
\rightarrow \infty$. If one of the radii in Eq.~\eqref{An} is infinite, then one obtains the interaction coefficient between an elastic plate and an elastic sphere: 
\begin{equation}
	A_{1,N+1} = \frac{4E_pE_{1,N}R_{1,N}^{1/2}}{3[E_{1,N}(1-\nu_p^2)+E_p(1-\nu_{1,N}^2)]}\,,
\end{equation}
where $E_p$ is the elastic modulus, $\nu_p$ is the Poisson ratio of the elastic plates at the boundaries, and the suffixes $1$ and $N+1$, respectively, indicate the left and right boundaries of the chain. Consequently, the equations of motion for spheres $1$ and $N$ are
\begin{align}
	\ddot{u}_1 &= \frac{A_1}{m_1}[\Delta_1-u_1]_+^{3/2}-
	\frac{A_{2}}{m_1}[\Delta_{2}+u_1-u_{2}]_+^{3/2}\,,\label{Hertz-boundaries1}\\
\ddot{u}_N &= \frac{A_N}{m_N}[\Delta_N+u_{N-1} -u_N]_+^{3/2} \notag \\ &\quad -
	\frac{A_{N+1}}{m_N}[\Delta_{N+1}+u_N]_+^{3/2}\,.
	\label{Hertz-boundaries2}
\end{align}

Equation~\eqref{Hertz} does not include effects from restitution or dissipation, so we assume that we can neglect energy that dissipates into internal degrees of freedom. Most investigations of granular crystals make these assumptions, and a conservative (and Hamiltonian) description of granular crystals has been extremely useful for numerous comparisons of theoretical and computational results to laboratory experiments \cite{Nesterenko:book,Sen:PR2008}, including in the presence of disorder 
\cite{Ponson:PRE2010,Fraternali:MAMS2010}. The proper physical form of dissipation is not known and is still a subject of ongoing debate. See Refs.~\cite{Sen:PR2008,nester2007,Carretero:PRL2009,vergara} for recent 
discussions of dissipative forces in granular crystals. 
Note additionally that we will not worry about incorporating
proper restitution forces, as we conduct our simulations in domains of sizes that ensure that the waves that we examine do not reach the domain boundary during the reported time horizon.

Even for homogeneous chains, Eq.~\eqref{Hertz} includes several interesting features that are not present in other lattice models (such as the well-known nonlinear Schr\"odinger (NLS) and Klein-Gordon (KG) lattices ~\cite{Flach:PR2008}). From a structural perspective, the present model is a type of FPU lattice~\cite{FPU}. It exhibits important differences from NLS and KG lattices, which typically include both a linear coupling and an on-site nonlinearity. However, there are also respects in which granular chains differ fundamentally from traditional FPU models \cite{Sen:PR2008}. In particular, when there is no precompression (which is known as the ``sonic-vacuum regime''~\cite{Nesterenko:book}), the sound
speed goes to $0$ and the system becomes purely nonlinear (i.e., linearizing it simply yields $0$). This allows solutions like compactons to occur in PDE limits of Eq.~\eqref{Hertz}. 
Additionally, because compactons are not exact solutions of the original granular chain (which has a fundamentally discrete nature), traveling waves in strongly nonlinear regimes exhibit a superexponential decay at the edge of the distribution instead of having compact support~\cite{chaterjee,Pikovsky:PhysicaD2006}. In the presence of precompression, which yields a linear term in Eq.~\eqref{Hertz}, Refs.~\cite{cho1,cho2} illustrated both analytically and numerically that energy-localizing states can arise in the form of dark breathers.


\subsection{Precompression Regimes}\label{regimes}

By changing the magnitude of the static load $F_0$ relative to the displacements $|u_n-u_{n+1}|$ between particles, one can tune the strength of the nonlinearity in Eq.~(\ref{Hertz}). To do this, we approximate the force using a power-series expansion, which is known to be suitable for a strongly compressed or weakly nonlinear chain \cite{Nesterenko:book}. We thereby distinguish three different regimes, which we now discuss.


\subsubsection{``Linear'' Regime ($\Delta_n\gg |u_{n-1}-u_n|$)}\label{regime1}

In this regime, we linearize Eq.~\eqref{Hertz} about the equilibrium state in
the presence of precompression to obtain
\begin{equation}
	m_n\ddot{u}_n = B_{n}u_{n-1}+B_{n+1}u_{n+1}-(B_n+B_{n+1})u_n\,,
\label{Linear}
\end{equation}
where
\begin{equation}\label{equ9}
	B_{n} =\frac{3}{2}A_n\Delta_n^{1/2} =  \frac{3}{2}A_n^{2/3}F_0^{1/3} \propto R^{1/3}.
\end{equation}
Note that we have neglected the higher-order terms (even the quadratic ones) in 
the expansion for very weak strains (i.e., small relative displacements). This linear limit corresponds to a chain of coupled harmonic oscillators. We represent the solutions to Eq.~\eqref{Linear} as complex wavefunctions to obtain a complete set of eigenfunctions of the form $u_n = v_n e^{i\omega t}$, where $\omega$ is
the eigenfrequency (so the eigenvalue is  $\lambda = -\omega^2$). 
In the homogeneous case --- i.e., for $B_n=B =
\text{const}$ for all $n$ --- we obtain plane waves $v_n = e^{ikn}$, and the dispersion relation, 
\begin{equation}\label{disperse}
	\omega = 2\pi f =\sqrt{\frac{2B}{m}\left[1-\cos(k)\right]}\,,
\end{equation}
where $m$ is the mass of a particle, gives a single acoustic branch.
The frequency satisfies the bounds $\omega \geq \omega_0 = 0$ and $\omega \leq \Omega =
\sqrt{4B/m}$, so the group velocity in this homogeneous case is
\begin{equation}
	v_g = \frac{\partial \omega}{\partial k} =
	\sqrt{\frac{B}{2m}}\frac{\sin(k)}{\sqrt{1-\cos(k)}}.
\end{equation}
The maximum of the group velocity is $v_g^m =\sqrt{B/m}=\Omega/2$, and it occurs when
$k=0$. From $v_g^m$, we are able to write expressions for several quantities. For instance, given an initially localized excitation $u_n(0) =\delta_{n,N/2}$, wave spreading takes place within a cone $\{N/2\pm v_g^m t,t\}$. 
Therefore, in our simulations, we consider systems that have at least $\lceil \Omega T\rceil$ spheres, 
where $T$ is the integration time and we recall that the ceiling function is $\lceil x\rceil = \text{min}\{k\in\mathds{Z}| k\geq x\}$.  This consideration allows us to avoid boundary effects
when we study dynamics. 

For an arbitrary arrangement of spheres in a granular chain, the eigenvalue problem associated with Eq.~(\ref{Linear}) takes the generic form
\begin{equation}
	\lambda {\bf v} = \Lambda {\bf v}\,, \label{eigeneq}
\end{equation}
where $\lambda = -\omega^2$ is the eigenvalue and $\Lambda = {\bf M}^{-1}{\bf B}$, where
${\bf M}_{ij} = m_{i}\delta_{i,j}$ are the elements of the diagonal matrix ${\bf M}$ of masses and 
\begin{equation}
	{\bf B}_{ij} =
B_{i+1}\delta_{i,j-1}+B_i\delta_{i,j+1}  - (B_i+B_{i+1})\delta_{i,j}
\end{equation}
are the elements of a tridiagonal symmetric matrix. Note that we have used fixed boundaries at both ends of the chain (see Eqs.~\eqref{Hertz-boundaries1} and \eqref{Hertz-boundaries2}). For a disordered chain (see Sec.~\ref{types} for the different types of disorder that we study), both matrices have random entries. Consequently, $\Lambda$ is an asymmetric tridiagonal matrix with random entries.


\subsubsection{``Weakly Nonlinear'' Regime ($\Delta_n > |u_{n-1}-u_{n}|$)}\label{regime2}

An intermediate regime between Eqs.~\eqref{Hertz} and~\eqref{Linear} is described by 
\begin{equation}
	m_n\ddot{u}_n = \sum_{i=1}^{3}\left[B_{n}^{(i)}(u_{n-1}-u_n)^i -
	B_{n+1}^{(i)}
	(u_n-u_{n+1})^i\right]\,,
	\label{HertzWNL}
\end{equation}
where
\begin{align}\label{b1b2b3}
	B_{n}^{(1)} &=B_n \propto R^{1/3}\,,\\
	B_n^{(2)} &= \frac{3}{8}A_n^{4/3}F_0^{-1/3}\propto R^{2/3}\,, \notag \\
	B_n^{(3)}&= -\frac{3}{48}A_n^2F_0^{-1}\propto R\,. \notag
\end{align}

This amounts to a particular case of the FPU model~\cite{FPU} that includes the so-called ``$\alpha$'' and ``$\beta$'' terms from two of the forms of nonlinearity in the original FPU model. 
One interesting feature of this regime is that small-amplitude intrinsic localized modes (ILMs, which are also often called ``discrete 
breathers'') \cite{Campbell:Today2004,Flach:PR2008} 
of the bright type (i.e., on top of a non-vanishing background) do not exist~\footnote{However, on top of a non-vanishing background, dark breathers can arise~\cite{cho1,cho2} from the linear limit.} in the absence of disorder because of the specific relations between the parameters $B_n^{(1)}$, $B_n^{(2)}$, and $B_n^{(3)}$ in Eq.~\eqref{b1b2b3}. This phenomenon was discussed in Ref.~\cite{Theocharis:PRE2009} based on the consideration of modulational instabilities (MIs) of linear waves due to nonlinearity.  An MI is a generic mechanism to generate such localized waves from linear waves at band edges of a linear spectrum.
However, to have an MI, it is necessary that $3B_n^{(1)}B_n^{(3)}-4B_n^{(2)}>0$, which is not satisfied in the present case.  Nevertheless, introducing impurities in a granular chain leads to the emergence of breather-like ``defect'' solutions that bifurcate from linear impurity modes~\cite{Theocharis:PRE2009}.


\subsubsection{``Strongly Nonlinear'' Regime ($\Delta_n\lesssim |u_{n-1}-u_n|$)}\label{regime3}

When precompression is sufficiently weak in comparison to the strains (and for vanishing precompression), one can no longer approximate Eq.~\eqref{Hertz} by truncating a Taylor expansion.  In general, for materials in which the sound speed goes to $0$ or remains very small, it is not reasonable to use a standard linear approximation as a starting point for a perturbative analysis~\cite{Nesterenko:book}. This is particularly interesting from the point of view of transport and localization theory in nonlinear disordered systems, because almost all of the research in the field has focused on the influence of nonlinearity for disordered systems in which the linear spectrum is initially either full of or partially full of localized states~\cite{Flach:ArxivRep2014-1,Flach:ArxivRep2014-2}. 
Consequently, understanding the interplay between disorder and nonlinearity in the sonic-vacuum limit brings new theoretical challenges, and --- as we shall see --- it also produces a fundamentally distinct form of dynamics. 

As a starting point towards developing a theory for transport and localization in granular crystals, we nevertheless start by focusing our efforts in a standard way by extending the linear theory to the nonlinear regime.


\begin{table}[th!]
\caption{\label{tab:tableratios}%
Calculations of the ratio $\epsilon = |u_{n_0 - 1}-u_{n_0}|/\Delta_{n_0}$ for an initial displacement excitation $\{u_n(0),\dot{u}_n(0)\}= \{\alpha\,\delta_{n,n_0},0\}$ in a homogeneous chain. 
We use the value $\alpha = 10^{-1}$ $\mu$m.}
\begin{ruledtabular}
\begin{tabular}{lc}
\textrm{$F_0$ (N)}&$\epsilon$\\
\colrule
$10$ &     $0.008$\\
 $1$ &     $0.362$\\
 $0.5$ &     $0.575$\\
 $0.1$ &    $1.682$\\
 $0.01$    &   $7.807$ \\
 $0$    &   $\infty$ \\
 \end{tabular}
\end{ruledtabular}
\label{tabla}
\end{table}

\subsection{Physical Parameters}\label{physical}

We take advantage of the numerous experimental investigations of granular crystals \cite{Sen:PR2008} to incorporate physically meaningful values for the parameters in Eq.~\eqref{Hertz}.  
We suppose that all the spheres are made of steel, and we use the parameters given in~\cite{Theocharis:PRE2009} unless we specify otherwise. In particular, the elastic modulus is $E =193$ GPa, 
the Poisson ratio is $\nu=0.3$, and the density is $8027.17$ kg/m$^3$. We also suppose that the elastic plates at the boundaries have the same mechanical properties as the 
spheres, so $\nu_1 = \nu_N = \nu$ and $E_1 = E_N = E$. In this paper, we examine disordered bidisperse granular chains, and we choose the radii of the spheres to be $R_1=4.76$ mm and $R_2=\xi R_1$, where $\xi\in(0,1]$. Note that $\xi = 1$ reduces the system to the case of a homogeneous chain. To explore different nonlinear regimes, we use $F_0$ in the range between $0$ N and $10$ N. 
As we discussed in Subsection \ref{regimes}, the amount of nonlinearity in the dynamics of each bead depends on the ratio 
$\epsilon_n = |u_{n - 1}-u_{n}|/\Delta_{n}$. In Table \ref{tab:tableratios}, we show the initial value of the ratio $\epsilon = \epsilon_{n_0}$ for a homogeneous chain with an initially localized displacement excitation $\{u_n(0),\dot{u}_n(0)\}= \{\alpha\,\delta_{n,n_0},0\}$ with $\alpha = 10^{-1}$ $\mu$m.


\section{Types of Disordered Configurations} \label{types}

\begin{figure}[th!]
\includegraphics[height=3.5cm]{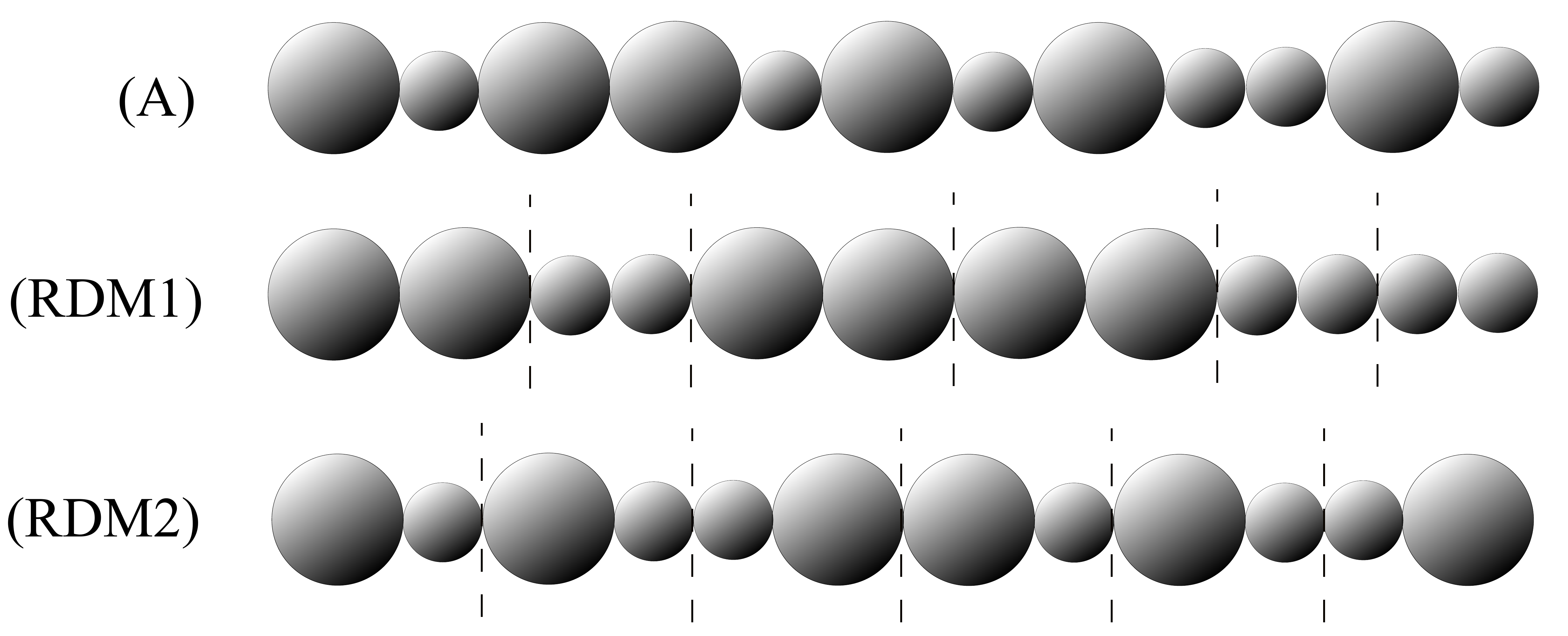}
\caption{Examples of different types of disordered chains. (A) Anderson model, (RDM1) random dimer model 1, and (RDM2) random dimer model 2. The vertical dashed lines are for visual guidance to    
separate adjacent dimers from each other.
}
\label{f5}
\end{figure}


In this article, we study three qualitatively different configurations of disordered granular chains: an Anderson-like configuration in which adjacent sites are uncorrelated and two types of random dimer models (RDMs) that include correlations across sites.

We consider bidisperse granular chains, so each chain consists of some configuration that includes two possible types of spheres: type 1 has radius  $R_1$, and type 2 has radius $R_2 = \xi R_1$, where $\xi\in (0,1]$. We also suppose that all of the spheres are made from the same material, so their elastic properties are the same. That is, $E_1=E_2=E$, $\nu_1 = \nu_2=\nu$, and the density is the same. We take their masses $m_1$ and $m_2$ to be different. Note that $\xi = 1$ reduces the system to the case of a homogeneous chain.
We consider three different ways of distributing the particles to produce disorder: (1) an Anderson-like distribution that amounts to uncorrelated disorder, (2) a random dimer distribution that follows a choice 
in Ref.~\cite{Dunlap:PRL1990} (RDM1), and a random dimer distribution that follows the choice in Ref.~
\cite{Ponson:PRE2010} (RDM2).
We show all three types of disorder in Fig.~\ref{f5}, and we note that both RDM1 and RDM2 are correlated types of disorder. See \cite{Izrailev2012125} for a review of localization in systems with correlated disorder.

Specifically, we construct our three families of disordered chains as follows:
\begin{enumerate}
\item{Anderson (A): For each of the $N$ particles in the chain, choose radius $R_1$ with a probability of $q \in [0,1]$ and radius $R_2$ with a probability of $1-q$. (Importantly, note that the choice for each particle is independent of all other particles.)}
\item{Random dimer model 1 (RDM1): For each of the $N/2$ dimers in the $N$-particle chain, we choose the configuration $R_1R_1$ (i.e., both particles have a radius of $R_1$) with a probability of $q \in [0,1]$ and the configuration $R_2R_2$ with a probability of $1-q$. (In the literature, this family specifically is what is usually meant by the term ``random dimer model'' \cite{Dunlap:PRL1990}.) 
}
\item{Random dimer model 2 (RDM2): For each of the $N/2$ dimers in the $N$-particle chain, we choose the configuration $R_1R_2$ with a probability of $q \in [0,1]$ and the configuration $R_2R_1$ with a probability of $1-q$.  
(That is, we are choosing the orientation of the dimer, which we imagine to be a spin with two possible states \cite{Ponson:PRE2010}.)
}
\end{enumerate}

Because these granular chains include two types of spheres (and are oriented horizontally, so we can ignore gravity), there are three types of sphere--sphere interactions: 
\begin{itemize}
\item{${A}_{11} =\frac{E\sqrt{2R_1}}{3(1-\nu^2)}$ (between two spheres of radius $R_1$),}
\item{${A}_{22} =\frac{E\sqrt{2R_2}}{3(1-\nu^2)}$ (between two spheres of radius
$R_2$),} 
\item{${A}_{12} =
\frac{2E}{3(1-\nu^2)}\left(\frac{R_{1}R_2}{R_{1}+R_2}
\right)^{1/2}$ (between spheres of different radii).}
\end{itemize}

One can characterize the disorder using two parameters. The parameter $q$ defines the extent of disorder. Thus, $q=0$ and $q=1$ are fully ordered cases, and $q=1/2$ is the
most disordered case \footnote{Note, however, that there are specific chain configurations --- such as a periodic sequence of dimers with alternating spin orientations in RDM2 --- for which $q = 1/2$ gives a maximal order with respect to higher-order correlations. Upon averaging over many configurations with the same value of $q$, such situations contribute little to the expected dynamics due to their low probability of occurrence.}. The other parameter is $\xi$, which is deterministic and defines the strength of the disorder by affecting the inertia (via the mass) of the particles and the magnitude of the interactions ${A}_{12}$ and ${A}_{22}$.


It is worth remarking that in the original Anderson configuration, the radius of the $n$th particle is $R_n = R_0 + \delta R_n$, where $\{\delta R_n\}$ corresponds to some uncorrelated sequence, such 
that $\delta R_n\in (-W,W)$ and $W$ is the
disorder strength. The ``Anderson'' model that we study (which is more precisely designated as ``Anderson-like'') is an example of a ``random binary alloy'' \cite{Evangelou:PRB1993} that has the same correlation properties as the original Anderson model. In the most general case, a random binary alloy can also include correlations due to dimer terms like the ones in RDM1 and RDM2 \cite{Evangelou:PRB1993}. We study the correlation properties 
for each type of disorder in the next subsection.


\subsection{Correlations}

Let ${\bf v_0}$ be a random vector generated by the rules that we described above.  
Without loss of generality, we label each entry $v_{0,i}$
of ${\bf v_0}$ as $0$ or $1$. Thus, the Anderson chain has vector components of
\begin{equation}
	v_{0,i} = \left\{\begin{array}{lcc}
	0\,,&\text{with probability}& q\\
	1\,,&\text{with probability}& (1-q)
	\end{array}\right.\,,
\end{equation}
where $i \in \{1,2,\ldots,N\}$. For the dimer models, we have 
\begin{equation}
	v_{0,i} = \left\{\begin{array}{lcc}
	c_0\,,&\text{with probability}& q\\
	c_1\,,&\text{with probability}& (1-q)
	\end{array}\right.\,,
\end{equation}
where $i \in \{1,2,\ldots, N/2\}$, $\{c_0,c_1\} = \{0\,0,1\,1\}$ for RDM1, and $\{c_0,c_1\} = \{0\,1,1\,0\}$ for RDM2.  

Let ${\bf v_n}$ be the $n$-cyclic permutation of ${\bf v_0}$ that satisfies
\begin{equation}
	v_{n,i} =\left\{\begin{array}{lcc}
	v_{0,i-n}\,,&\quad\text{if}& i>n\,,\\
	v_{0,N-n+i}\,,&\quad\text{if}& i\leq n\,.
	\end{array}\right.
\end{equation}
To characterize the amount of correlation in each case, we calculate
the Pearson correlation coefficient 
\begin{equation}
	\rho_{n,n'} = \frac{\text{cov}
	\left({\bf v_n},{\bf v_{n'}}\right)}{\sigma_{{\bf v_n}}\sigma_{{\bf v_{n'}}}}\,,\label{Pearson}
\end{equation}
where $\text{cov}({\bf v_n},{\bf v_{v'}}) =
\text{E}\left[({\bf v_n}-\bar{v}_n{\bf I})({\bf v_{n'}}-\bar{v}_{n'}{\bf I})\right]$
is the covariance between ${\bf v_n}$ and ${\bf v_{n'}}$, the vector ${\bf I} = (1,1,\ldots,1)$ has all elements equal to $1$, the standard deviation of the vector ${\bf v}$ is $\sigma_{{\bf v}}=\sqrt{\text{E}\left[({\bf v}-\bar{v}{\bf I})^2\right]}$, and the mean of ${\bf v}$ is given by the expectation $\text{E}\left[{\bf v}\right] = \bar{v}$. For our computations, it is convenient to write the covariance as
\begin{align}
	\text{cov}\left({\bf v_n},{\bf v_{n'}}\right) &= \frac{1}{N}\sum_{i=1}^N 
	\left[v_{n,i}v_{n',i}-(v_{n,i}\bar{v}_{n'}+v_{n',i}\bar{v}_n)\right.\nonumber\\&\hspace{2cm}
	\left.+\bar{v}_n\bar{v}_{n'}\right]\,.
\label{cov}
\end{align}

Note that some statistical properties, such as the mean and
standard deviation, are independent of permutations (i.e., $\bar{v}=\bar{v}_n=\bar{v}_{n'}$ and $\sigma_{\bf v}=\sigma_{\bf v_n}=\sigma_{\bf v_{n'}}$), so we can write Eq.~\eqref{Pearson} as
\begin{equation}
	\rho_{n,n'} = \frac{\left(\frac{1}{N}\sum_{i=1}^N
v_{n,i}v_{n',i}\right)-\bar{v}^2}{\sigma_{\bf v}^2}\,,
\end{equation}
where one can calculate the term in parentheses as the sum of conditional
probabilities that depend on the type of disorder. 

In the next three subsubsections, we calculate the correlation coefficients for each type of disorder in the thermodynamic (i.e., $N\rightarrow \infty$) limit.

\begin{figure}[t]
\includegraphics[height=5.5cm]{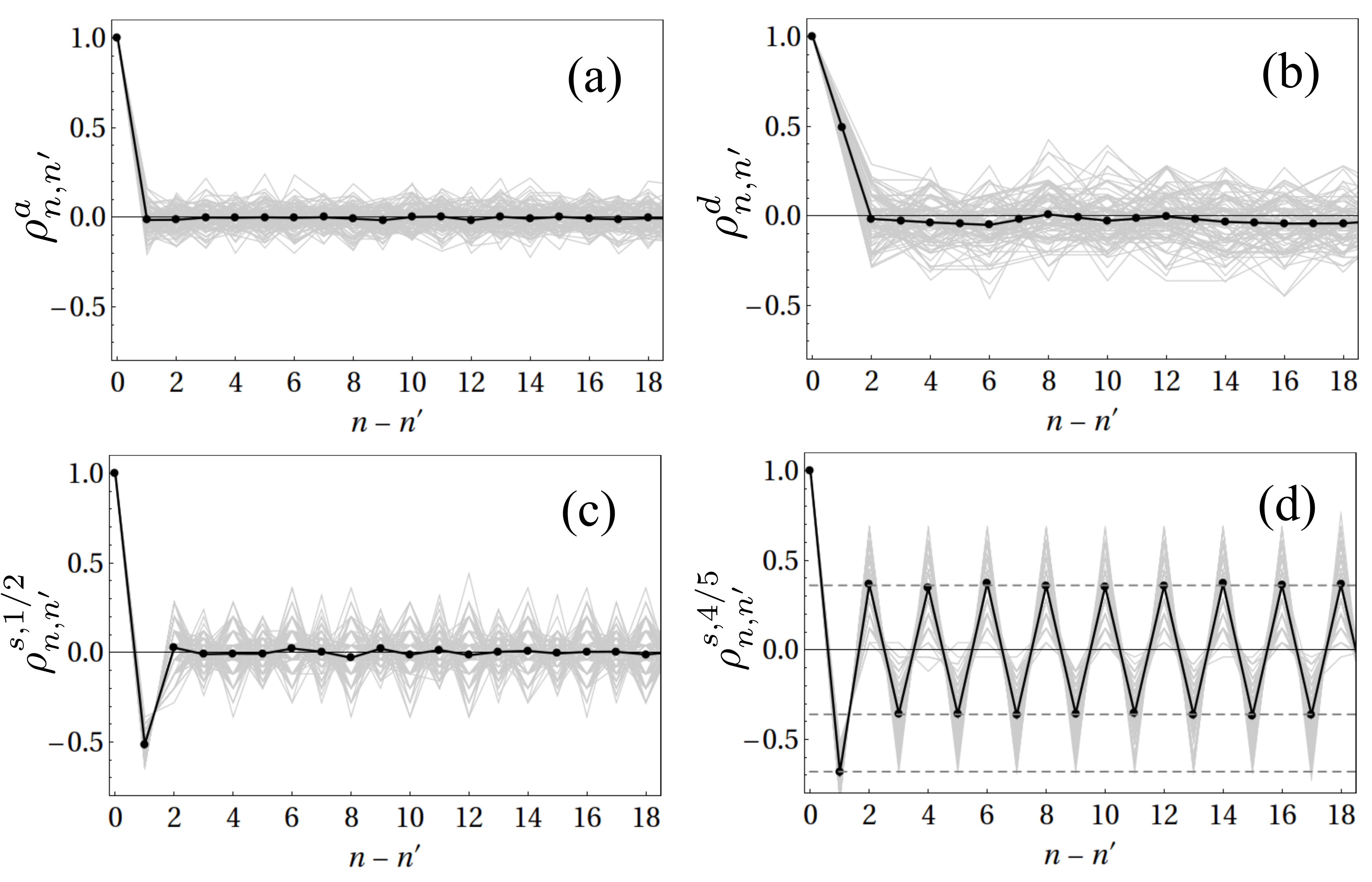}
\caption{Correlation function $\rho_{n,n'}$ versus the distance between particles for the three types of disordered granular chains: 
(a) Anderson model, (b) random dimer model 1 (RDM1), and (c,d) random dimer model 2 (RDM2). The gray curves show 100 realizations for a 
chain with $N=100$ particles, and the black curves give the mean values. In panels (a)--(c), we use $q=1/2$; in panel (d), we use $q=4/5$. 
The horizontal dashed lines in panel (d) show analytical values for the long-range correlation from Eq.~\eqref{spincorr}.}
\label{f-cor}
\end{figure}

\subsubsection{Anderson}

In a granular chain with an Anderson-like disorder, the mean value of ${\bf v}$ is $\bar{v} = (1-q)$, and the
standard deviation is $\sigma_{\bf v} = \sqrt{q(1-q)}$. Both quantities
depend on the probability $q$, but the correlation
\begin{equation}\label{and-uncorr}
	\rho_{n,n'}^a = \delta_{0,|n-n'|}
\end{equation}
is independent of $q \in (0,1)$. When $q=0$ or $q=1$, the correlation becomes $\rho_{n,n'}^a = 1$ because the granular chain is homogeneous. Equation \eqref{and-uncorr} 
implies that the Anderson disorder is an uncorrelated type of disorder. This is true exactly in the thermodynamic limit (i.e., as the number of particles $N \rightarrow \infty$). However, it is also true 
in an average sense for finite systems, which implies that the mean of the Pearson correlations for a large number $S\rightarrow \infty$ of finite systems approaches the value of the correlation for a single 
system as $N\rightarrow \infty$. In Fig.~\ref{f-cor}(a), we show the Pearson correlation coefficient as a function of the relative distance between spheres. The black curve shows the mean value, which tends to $\delta_{0,|n-n'|}$ (as we just discussed).


\subsubsection{Random Dimer Model 1 (RDM1)}\label{rdm1-sec}

As in the Anderson case, the mean value of ${\bf v}$ for the RDM1 granular chain is $\bar{v} = (1-q)$,
and the standard deviation is $\sigma_{\bf v} = \sqrt{q(1-q)}$. However, because an RDM1 granular chain consists of a sequence of dimers, there is now a short-range correlation.  
The Pearson correlation coefficient is
\begin{equation}
	\rho_{n,n'}^{d}= \delta_{0,|n-n'|}+
	\frac{1}{2}\delta_{1,|n-n'|}\,,
\end{equation}
which we note is again independent of $q \in (0,1)$. Consequently,
the chain has the above short-range correlation between second-nearest-neighbors
neighbors for any $q\in (0,1)$.

In Fig.~\ref{f-cor}(b), we show the Pearson correlation coefficient as function of the relative
distance between spheres.


\subsubsection{Random Dimer Model 2 (RDM2)}

The RDM2 granular chain has rather different statistical properties from the other two types of disordered chains.

The mean value of ${\bf v}$ is $\bar{v} = 1/2$, and the standard deviation is $\sigma_{\bf v} =
1/4$. Both the mean and the standard deviation are independent of the probability $q$, because $q$ affects only the orientation of the the dimer; the numbers of $0$ values and $1$ values are 
unchanged. This type of disorder includes a long-range correlation that one can tune with the parameter $q$.  The Pearson correlation coefficient is
\begin{align}
	\rho_{n,n'}^{s,q} &= \delta_{0,|n-n'|}-
	\frac{1}{2}\left[(2q-1)^2+1\right]\delta_{1,|n-n'|}\nonumber\\&\hspace{1cm} + \sum_{j=2}^N
	(-1)^j(2q-1)^2\delta_{j,|n-n'|}\,.
\label{spincorr}
\end{align}
An interesting special case occurs when $q=1/2$, as the correlation reduces to a short-range
anti-correlation:
\begin{equation}
	\rho_{n,n'}^{s,1/2} = \delta_{0,|n-n'|}-\frac{1}{2}\delta_{1,|n-n'|}\,.
\end{equation}

Other interesting limits are the ordered diatomic chains that arise for
$q = 0$ and $q=1$.  Because the orientation of the dimer units is constant in these limits, there is a perfect correlation between particles that are an even distance apart and a perfect anti-correlation between particles that are an odd distance apart:
\begin{equation}
	\rho_{n,n'}^{s,0} =\rho_{n,n'}^{s,1}= \sum_{s=0}^N (-1)^s\delta_{s,|n-n'|}\,.
\end{equation}

In Figs.~\ref{f-cor}(c,d), we show the Pearson correlation coefficient as a function of the relative
distance between spheres. We use different values of $q$ for the two panels.


\section{Impurities in a Homogeneous Granular Chain}\label{Sec4}

Reference~\cite{Theocharis:PRE2009} confirmed for granular
chains the general notion that either localized or resonant modes arise when an otherwise homogeneous system (a so-called ``host'' chain) includes impurities. The nature of such modes depends on the
relation between the parameters of the impurities and those of the other spheres
in the host chain. If an impurity mass is smaller (respectively, larger)
than the rest of the particles, then the associated mode
is localized (respectively, resonant). Moreover, it is possible to extend localized linear modes into the weakly nonlinear regime using a continuation procedure. 

Impurities also break the translational symmetry of a chain, which implies that scattering processes around
the impurities play a significant role in the dynamics. This becomes increasingly important as the number of impurities in a chain increases. To emphasize the role of impurities in the transport
and localization properties of a system, we highlight the so-called ``random dimer model'' \cite{Dunlap:PRL1990}, which we call RDM1 in the present article. In Ref.~\cite{Dunlap:PRL1990}, it was shown for the Schr\"odinger lattice with on-site energy distributed in an RDM1 way that --- even when almost all of the linear modes are spatially localized --- there is always one mode that is extended for a certain value (which depends on the strength of the disorder) of the wavenumber $k_d$. Furthermore, for finite 1D systems, there is a set of modes for wavenumbers near $k_d$ (in particular, for wavenumbers $k \in (k_d-\Delta k,k_d+\Delta k)$, with $\Delta k \sim1/\sqrt{N}$ as $N \rightarrow \infty$) that have a localization length that is larger than the length of the system \cite{Datta:PRB1995}. 

 A similar effect from double impurities has been observed in acoustic chains with harmonic interactions \cite{Datta:PRB1995}. However, due to the acoustic 
characteristics of the linear spectrum, the $0$-frequency linear mode is extended in either a homogeneous or a disordered system (independently of the type of disorder) and for any system length. 
Therefore, even for an Anderson-like disorder configuration, modes with $k\in [0,\Delta k)$ and $\Delta k\sim 1/\sqrt{N}$ (as $N \rightarrow \infty$) have a localization length that is larger than the size of the system
 \cite{Datta:PRB1995,Lepri:PRE2010}. We expect reflectionless modes to emerge in strongly 
compressed granular chains --- i.e., in the linear regime (see Sec.~\ref{regime1}), in which a harmonic approximation of the interactions is suitable.


\begin{figure*}[th!]
\includegraphics[height=12cm]{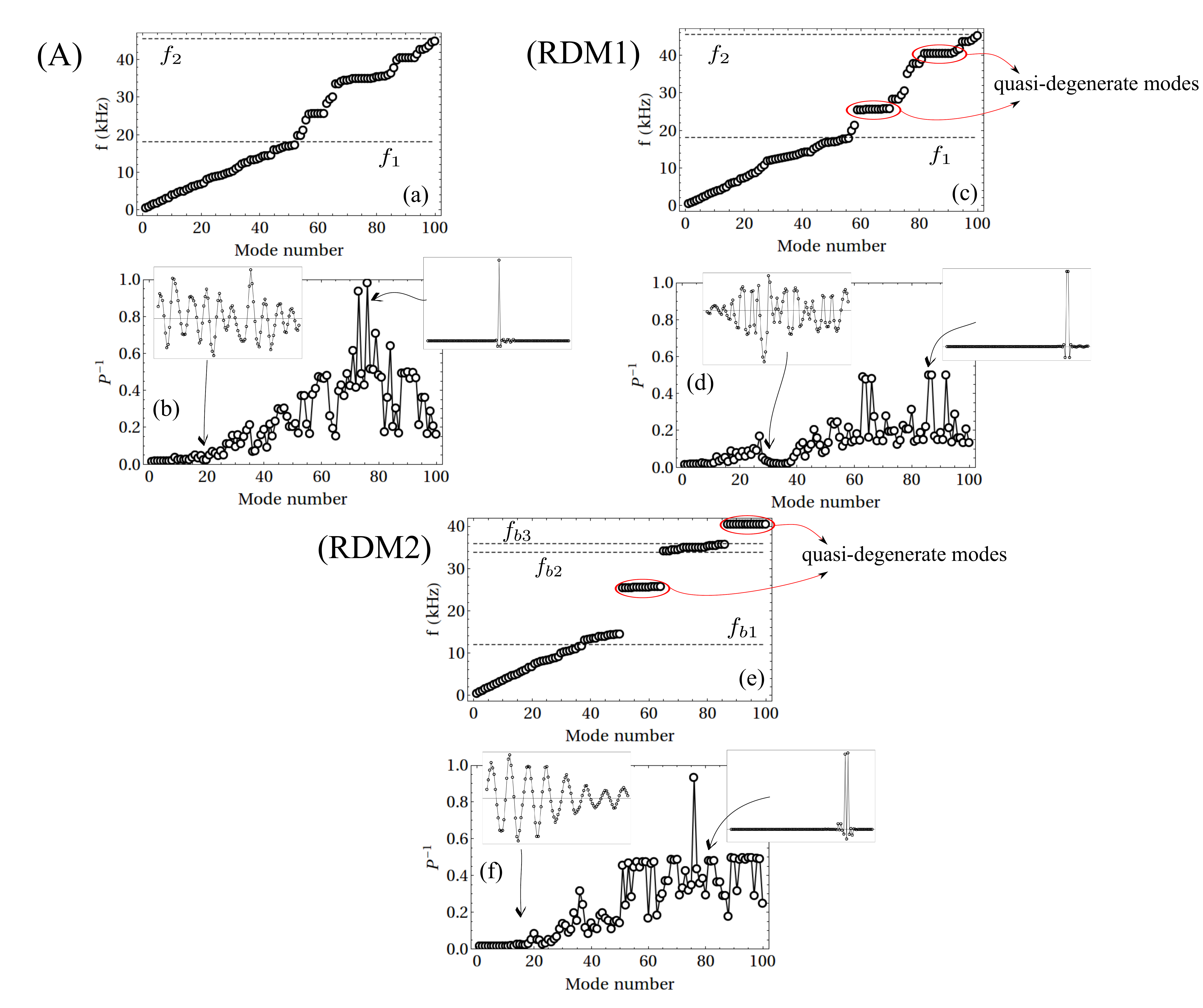}
\caption{Linear spectrum and inverse participation ratio $P^{-1}$ for different types of disorder
with probability parameter $q=0.5$ and size parameter $\xi = 0.5$. As usual, (A), (RDM1), and (RDM2) denote the Anderson, random dimer model 1, and random dimer model 2, respectively. 
The insets show examples of linear modes for both high and low frequencies. The dashed lines mark the cutoff frequency $f_i =
\frac{1}{2\pi}\sqrt{\frac{4B_{ii}}{m_i}}$ associated with a homogeneous
chain and the cutoff frequency $f_{bi} =
\frac{1}{2\pi}\sqrt{\frac{2B_{12}}{m_i}}$ for a diatomic chain, where $m_3 = m_1m_2/(m_1+m_2)$. We have also considered $B_{ij} =\frac{3}{2}A_{ij}\Delta_n^{1/2}$, with $A_{ij}$ described in Sec.~\ref{types}.
Using the parameter values described in Sec.~\ref{physical} and $F_0 = 10$ N,
we obtain $f_1 \approx 18.09$ kHz, $f_2 \approx 45.58$ kHz,
$f_{b1}\approx 11.95$ kHz,
$f_{b2}\approx 33.81$ kHz, and $f_{b3}\approx 35.86$ kHz.
}
\label{f8:linearspectrum}
\end{figure*}

\section{Numerical Results}\label{Sec5}

In general, it is difficult to precisely determine localization properties in disordered systems --- primarily because most tests are based on the asymptotic behavior of particular observables
(e.g., energy). From a practical perspective, one needs to consider long chains (and large volumes in larger dimensions) and very long integration times, and (from a theoretical perspective) one should let both time and system size go to infinity~\cite{Kramer:RPP1993}. Such scenarios are difficult to achieve experimentally, and even numerical simulations pose considerable difficulties \cite{Skokos:PRE2009}. In particular, one is often interested in the asymptotic behavior of the energy distribution. Hence, to conduct long-time simulations without significant (and unphysical) variation in a system's total energy, it is necessary to employ carefully-chosen numerical-integration schemes. Additionally, because we are examining disordered systems and we thus need to average over a large number of realizations of a particular type of disorder to obtain appropriate statistical power, it is also necessary to employ sufficiently fast numerical-integration schemes that are also particularly accurate in their energy conservation. We thus use a symplectic integrator from Refs.~\cite{Laskar:Celest2001,Skokos:PRE2009,skokos-erratum}.

We also rely on indirect methods to develop intuition about the asymptotic behavior of disordered granular chains. One such method is to study the structure of the linear spectrum and the extent of 
localization of the linear modes. For instance, in the classical Anderson model in a 1D electronic system~\cite{Anderson:PR1958}, all of the linear modes are localized exponentially for any amount of 
disorder. This leads to an absence of diffusion that manifests as a saturation of the second moment of the probability distribution as a function of time. In other words, excitations remain spatially 
localized.  By contrast, as we mentioned in Sec.~\ref{Sec4}, the RDM1~\cite{Dunlap:PRL1990} behaves differently from the Anderson model in this respect, as the former includes extended modes that cause the second moment to grow as a function of time. 

In our ensuing discussions, we investigate the influence of the three different types of disorder on the structure of the linear spectrum and the presence of localized states in both the bulk and the surface of a granular chain.  We subsequently investigate transport and dynamical localization in the bulk for disordered Hertzian chains \eqref{Hertz}.

\begin{figure*}[th!]
\includegraphics[height=12cm]{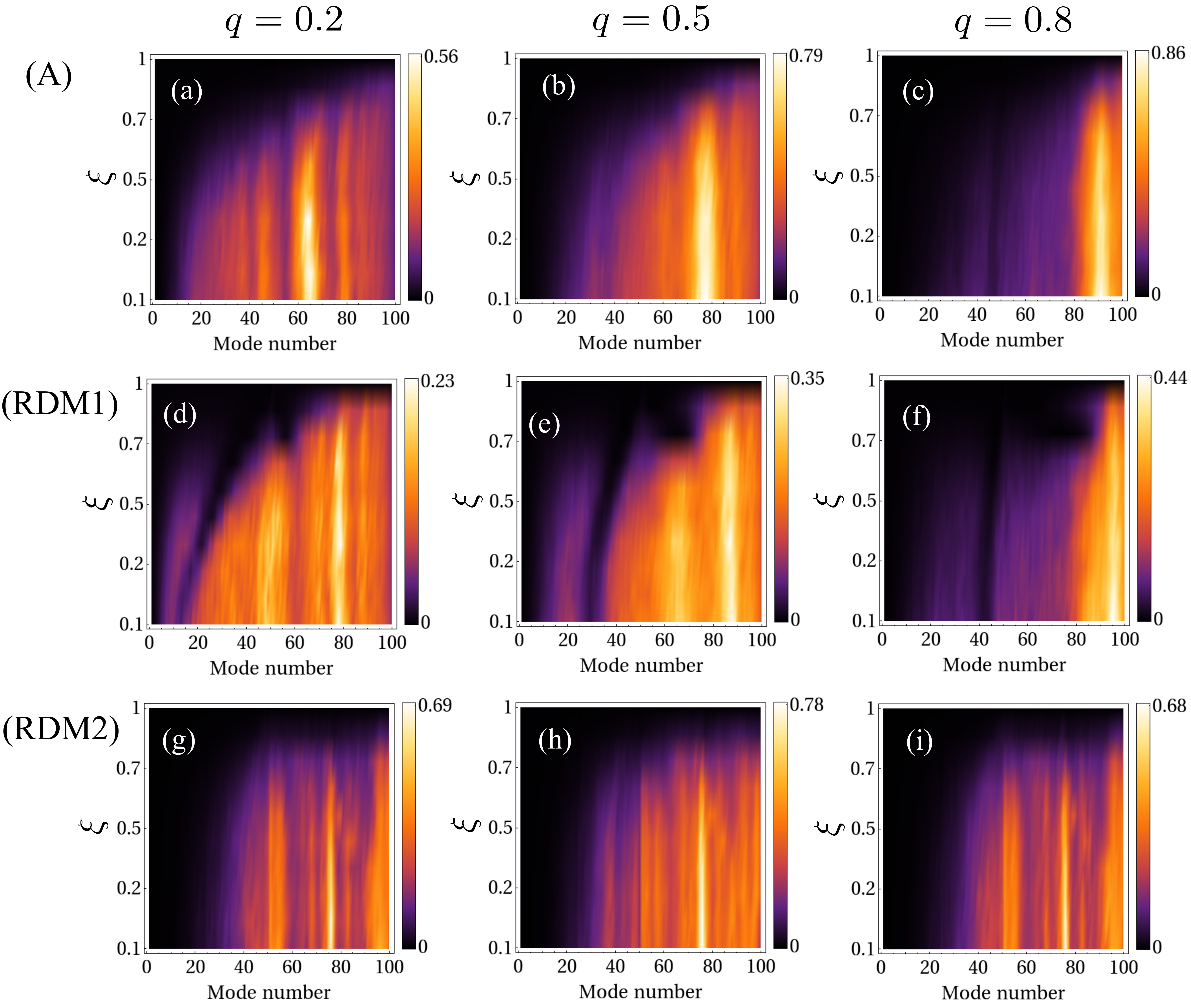}
\caption{Inverse participation ratio (averaged over 100
chain configurations) as a function of the ratio $\xi = R_2/R_1$
of radii and the mode number for different types of disordered chains and different
values of the probability parameters $q$. The black regions are associated with delocalized
waves.
} 
\label{f9:iprmap}
\end{figure*}

\subsection{Direct Diagonalization of Eq.~\eqref{eigeneq}}\label{numericalspectrum}

There are various ways of measuring localization in linear modes.
In finite systems, it is useful to calculate the inverse participation ratio (IPR) \cite{Kramer:RPP1993}
\begin{equation}
	P^{-1} = \frac{\sum_{n=1}^N h(v_n,\dot{v}_n)^2}{\left(\sum_{n=1}^N h(v_n,\dot{v}_n)\right)^2}\,,\label{IPReq}
\end{equation}
where $h(v_n,\dot{v}_n)$ some certain distribution. For modal analysis, we use $h(v_n,\dot{v}_n) = v_n^2$,
which allows one to measure the fraction of particles whose displacement of position from equilibrium
differs markedly from $0$. We can thereby measure the extent of localization. For instance, a plane wave with all sites equally excited satisfies $P^{-1}\rightarrow 0$ as the number of 
particles $N \rightarrow \infty$. By contrast, a strongly localized wave satisfies 
$P^{-1}\rightarrow 1$, and $P^{-1} =  1$ exactly when only one sphere is vibrating (i.e., when $v_n=0$ for all $n\neq j$ and $v_j\neq 0$).

Calculating the IPR makes it possible to directly obtain a qualitative understanding of the nature of the linear modes. In Fig.~\ref{f8:linearspectrum}, we show the 
spectrum and the extent of localization (i.e., its IPR) associated with the linear modes 
for one realization of each of the three types of disorder. In Fig.~\ref{f9:iprmap}, we show the mean value of the IPR over 100 realizations of each type of disordered chain as 
a function of the probability parameter $q$ and the size parameter $\xi$. In both figures, we have sorted the modes from smallest frequency to largest frequency.
Diagonalizing Eq.~\eqref{eigeneq} directly yields the displacement distribution of the particles in the chain that are associated with the different modes. In this section, we use these displacement
distributions to compute the IPRs that we show in Figs.~\ref{f8:linearspectrum} and~\ref{f9:iprmap}. We also evaluate Eq.~\eqref{IPReq} using the energy-density distribution (given by Eq.~\eqref{energy}, as we will discuss in Sec.~\ref{m2-energies}), and we obtain qualitatively similar results.  For each type of disorder, we will use the energy-density distribution (see Sec.~\ref{diff-nonlin}) to characterize the dynamical localization.

We first consider Anderson-like disorder. For frequencies $f \in (f_1,f_2)$ [see Eq.~\eqref{disperse} and Fig.~\ref{f8:linearspectrum}], we observe a complicated gap structure that includes isolated frequencies between the two band-edge frequencies. In the frequency range $(f_1,f_2)$, there is also a small region in which $P^{-1}$ has multiple peaks with values that are close to $1$. These peaks are associated with single-node impurity-like modes, in which the energy oscillates primarily around one particle. 
As was discussed in Ref.~\cite{Theocharis:PRE2009}, linear localized modes are bound to small particles for a single impurity, and the frequency $f_{\mathrm{imp}}$ of these modes is larger than the lower edge frequency $f_1$ of the homogeneous chain. Additionally, for a given precompression force $F_0$, the frequency $f_{\mathrm{imp}}$ depends only on the strength of the impurity, and it thus depends only on the size parameter $\xi$. There are also modes with $P^{-1} \approx 0.5$ that are related to double impurities. More precisely, $P^{-1}$ is slightly \emph{smaller} than $0.5$ because the mode does not consist exactly of two particles that vibrate, as there is also 
a tail that decays as a function of space. Modes with a lower IPR are associated with different local configurations. For example, a mode with two small masses that vibrate with a large amplitude and are 
separated by a large mass that oscillates with a small amplitude has $P^{-1}\approx 0.4$. 
Additionally, modes that have $5$ spheres that effectively 
participate in the system dynamics, for $N = 100$, have $P^{-1}\approx 0.2$, and one can make analogous statements for other values of $P^{-1}$.

One can interpret the probability parameter $q$ as a measure of the density of small impurities (i.e., particles with radius $R_2$) in a host chain of particles with radius $R_1$. As $q\rightarrow 1$, the granular chain is composed almost exclusively of spheres with radius $R_1$, and its few small impurities generate impurity modes whose 
frequencies are larger than $f_1$. The rest of the spectrum consists mostly of an acoustic branch that is bounded above by $f_1$. This explains why the Anderson chain with $q=0.8$ in Fig.~\ref{f9:iprmap} has an 
IPR whose maximum occurs near the maximum mode number (i.e., it is close to the frequency edge $f_1$). When $q$ decreases, the fraction of particles with radius 
$R_2$ increases, and the population of modes with frequencies between $f_1$ and $f_2$ increases as well. In particular, the maximum value of $P^{-1}$ in 
Fig.~\ref{f9:iprmap} in the Anderson case (which occurs for $q=0.2$) is about $0.55$, which implies that most localized linear modes are double impurity-like 
modes instead of single impurity-like modes. However, the frequency of these modes does not change for a fixed value of $\xi$, and it is close to the frequency edge at $f_1$.

Another interesting feature of the Anderson model in granular chains is that the $0$-frequency mode is extended for all values of $q$ and $\xi$. In other words, it is independent of the amount 
of disorder and of the relative sizes of the two types of 
particles \cite{Datta:PRB1995,Lepri1}. Near $\omega=k=0$, there is a nontrivial region 
in the $\xi$-$q$ parameter space in which one observes extended modes in a finite-size chain.  
One expects the area of this region to vanish as the system size $N \rightarrow \infty$~\cite{Datta:PRB1995}. However, the presence of this extended mode opens a channel for the transportation 
of energy even in a disordered chain.

For an RDM1 chain, the frequency structure is similar to that of an Anderson chain. However, there are several high-frequency modes, which each have frequency between $f_1$ and $f_2$, that form an almost flat structure 
in plots of frequency versus mode number (see Fig.~\ref{f8:linearspectrum}). These frequencies are related to {\it quasi-degenerate} modes, which have almost the same frequency as each other, and 
such modes arise more often in an RDM1 chain than in an Anderson-like chain. As in the Anderson-like chain, an RDM1 chain also includes some highly localized linear modes that are related to double impurities. Nevertheless, the main difference arises in the $P^{-1}$ distribution, which for an RDM1 chain includes an 
extra minimum near a frequency of $f_b\in(0,f_1)$ that depends on the parameters $\xi$ and $q$. For example, when $q=\xi=0.5$, we obtain roughly $f_b\approx 15$ kHz for $N = 100$ and the physical parameters described 
in Sec.~\ref{physical}. This is related to extended modes that are centered at a {\it nonzero} frequency. Furthermore, as one can see from Fig.~\ref{f9:iprmap}, the IPR tends to be smaller for most values of $\xi$ and $q$ 
in an RDM1 chain in comparison with an Anderson case. This occurs because the impurities in RDM1 chains are twice as large as those in Anderson chains, which implies in turn that RDM1 chains have large impurity modes.

An RDM2 chain exhibits completely different --- and rather remarkable --- features in its spectrum and IPR distribution from the other two types of disordered chains. To explain these 
differences, it is important to interpret the RDM2 system as a perturbation of a perfectly ordered diatomic chain 
instead of as a perturbation of a monoatomic one. In fact, most of the 
eigenvalues for an RDM2 chain occur between the frequency edges of the ordered diatomic chain (i.e., within its
pass bands). The rest of the eigenvalues are organized predominantly into almost flat distributions within the band gaps (see Fig.~\ref{f8:linearspectrum}). An RDM2 chain tends to have more degenerate 
modes than an RDM1 chain. RDM2 chains also have very interesting localization properties. In Fig.~\ref{f9:iprmap}, for example, we observe that the $P^{-1}$ distributions are (on average) 
almost independent of the degree of disorder
(i.e., on the parameter $q$). We also see from Fig.~\ref{f8:linearspectrum} that most of the degenerate modes are also equally localized. In other words, they have almost the same 
value of $P^{-1}$. To explain the features of the IPR, observe that there exist a few single impurity-like modes with $P^{-1}\approx 1$, but most of the localized modes consist of two 
(associated with $P^{-1}\approx 0.5$), three ($P^{-1}\approx 0.33$) or four ($P^{-1}\approx 0.25$)
vibrating particles. Additionally, the RDM2 disorder is symmetric with respect to $q=0.5$ by construction (so, e.g., $q=0.2$ and $q=0.8$ are equivalent situations).


\begin{figure}[th!]
\includegraphics[height=16cm]{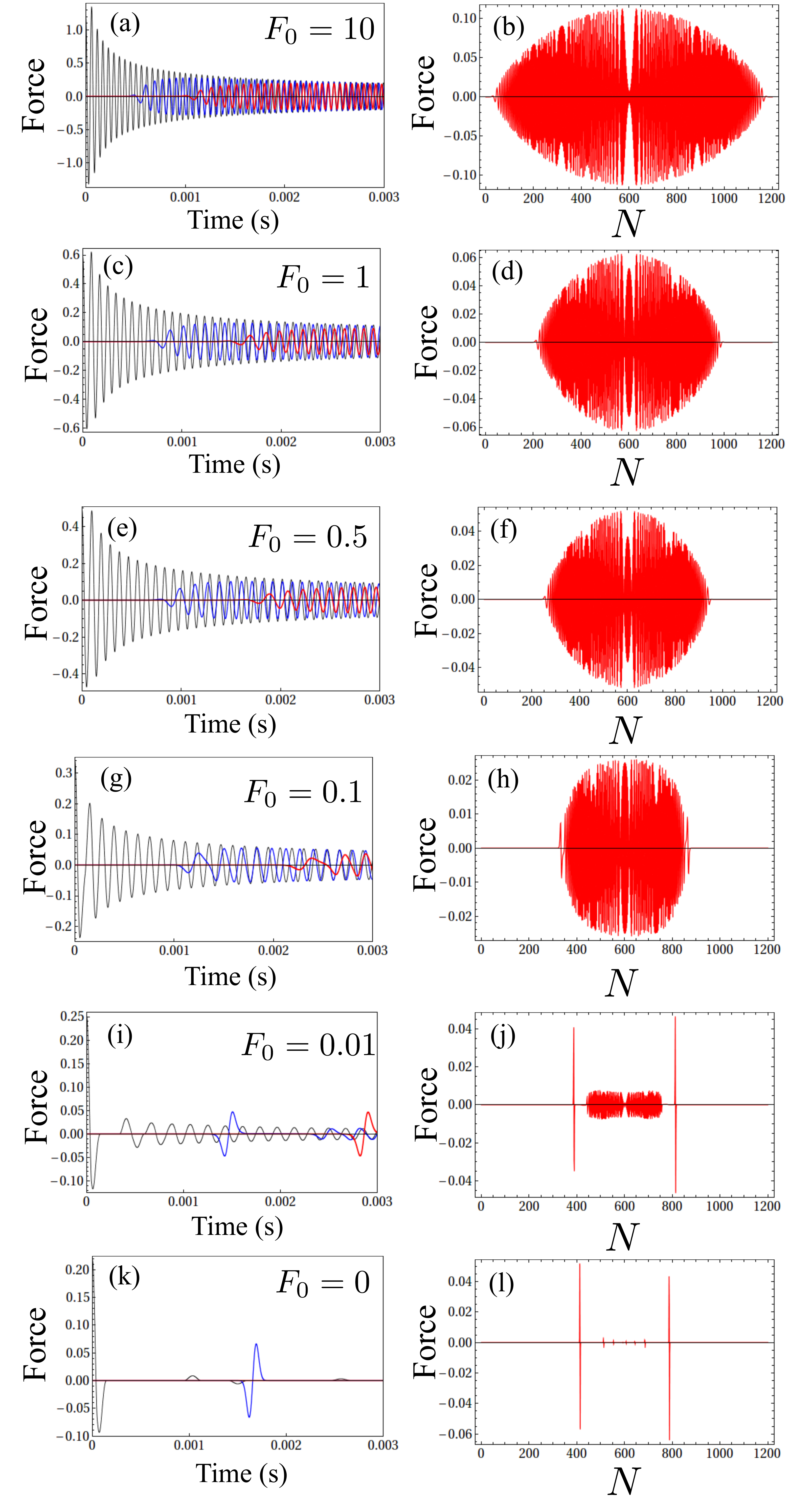}
\caption{(Left) On-site force as a function of time and (right) force distribution of particles for various amounts of precompression when we apply an excitation that consists of an initially localized displacement to the center of a homogeneous granular chain. In the left panels, the black curves give the force for particle 601, the blue curves give the force for particle 631, 
and the red curves give the force for particle 661. The chain has $N=1201$ particles.   For each example, the initial condition is 
$u_n=10^{-1}\times \delta_{n,601}$ $\mu$m. For the right panels, we give the force in Newtons at time $t=10^{-2}$ s.
In each row, the two panels are both for a chain with the same specified precompression strength.}
\label{f2-profiles}
\end{figure}

\begin{figure}[th!]
\includegraphics[height=16cm]{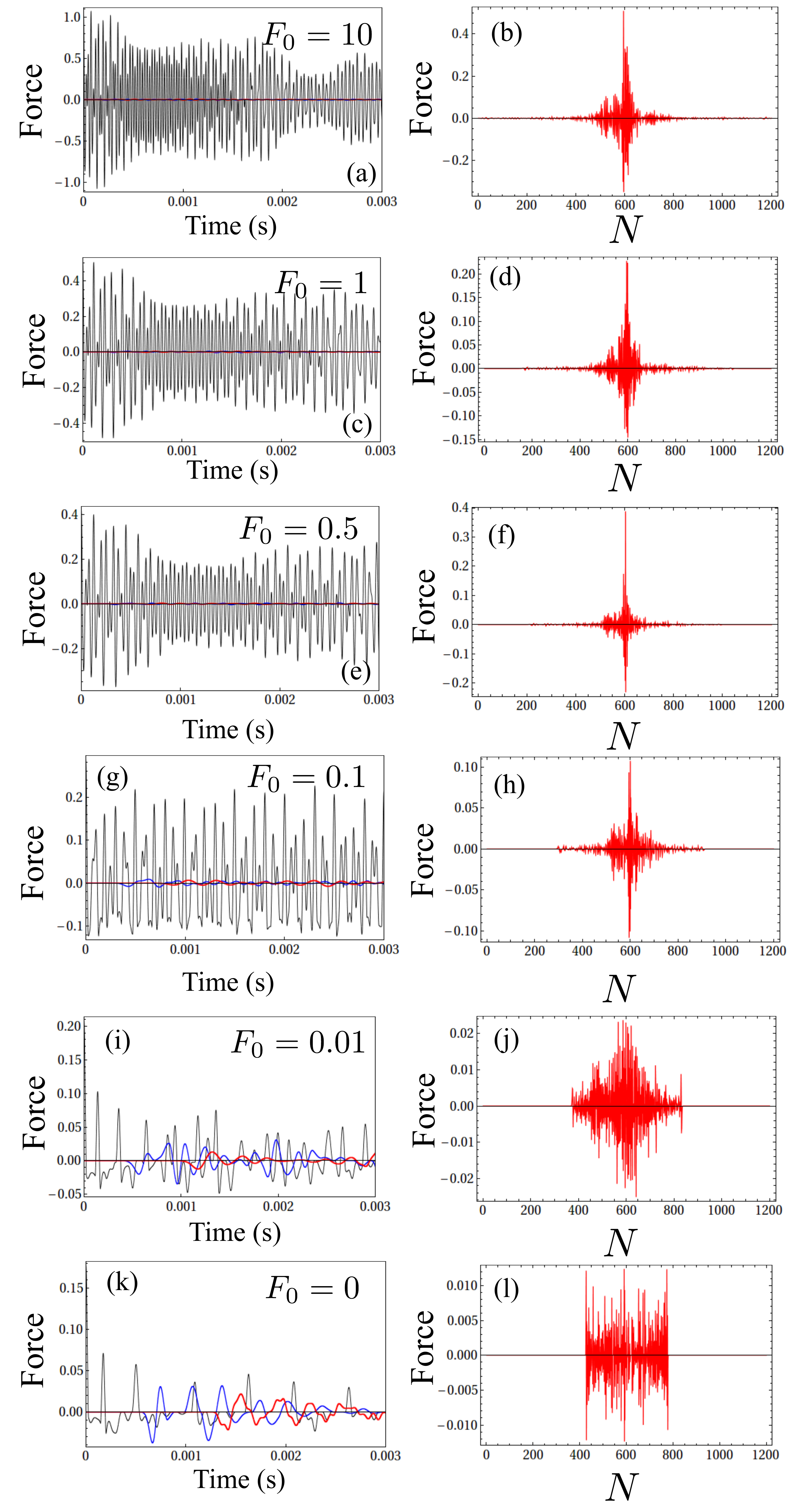}
\caption{(Left) On-site force as a function of time and (right) force distribution of particles for various amounts of precompression when we apply an excitation that consists of an 
initially localized displacement to the center of an Anderson chain with a particle-size parameter of $\xi=0.8$.
In the left panels, the black curves give the force for particle 601, the blue curves give the force for particle 631, and the red curves give the force for particle 661.
The chain has $N=1201$ particles.   For each example, the initial condition is 
$u_n=10^{-1}\times \delta_{n,601}$ $\mu$m. For the right panels, we give the force in Newtons at time $t=10^{-2}$ s. 
In each row, the two panels are both for a chain with the same specified precompression strength.
}
\label{f11-profilesAnderson}
\end{figure}


\subsection{Spreading and Partial Localization Due to Disorder and Nonlinearity}

Force distributions are particular useful in granular crystals, because it is easier and more reliable to measure forces than to measure energy. Moreover, examining forces as a function of time allows one to indirectly measure spreading and localization. 
Thus, in this section, we examine how the force evolves at specific spots in the chain and also how the force distributions are affected by changes in the precompression in homogeneous and Anderson-like disordered chains.

In Fig.~\ref{f2-profiles}, we show example force distributions from applying an excitation that consists of an initially localized displacement at the center of a homogeneous chain. For a strongly 
compressed chain (e.g., for $F_0=10$ N), the initial excitation spreads along the chain, and the dynamics arise from the decomposition of the Kronecker $\delta$ into linear modes. However, the spreading is slightly asymmetric, because the nonlinearity cannot be neglected entirely. Increasing the nonlinearity in the 
system by decreasing the precompression leads to a lessening of the distribution width due a decrease in the system's sound speed. One directly observes this effect in the force distribution, and one 
can also see it indirectly by examining the force at different places in the chain as a function of time. For instance, the time that takes to detect fluctuations in the force at particles 30 and 60 
sites away from the position of the initial
excitation becomes longer as one decreases the precompression. Additionally, in the sonic-vacuum regime, solitary waves emerge clearly, and the energy is divided mainly into two pulses that move in 
opposite directions. 

In the presence of disorder (see Fig.~\ref{f11-profilesAnderson}), we observe that the spatial force distribution changes abruptly (i.e., even for a small amount of disorder) from the distribution in an associated homogeneous chain. 
When linear effects are dominant (e.g., at $F_0=10$ N), the force distribution has a maximum near the position of the initial excitation, and it decays exponentially away from this point. Near the central position of the distribution, the temporal force dynamics includes large-amplitude, persistent oscillations that exist for long times. 
The forces in particles that are a few sites away from the center (e.g., see the particles that are $30$ and $60$ sites away from the center in Fig.~\ref{f11-profilesAnderson}) exhibit 
oscillations whose amplitudes are orders-of-magnitude lower. 

When we increase the effective nonlinearity in a granular chain --- in particular, 
in the weakly nonlinear situation, such as the one in Eq.~\eqref{HertzWNL} --- resonances of linear modes are induced by nonlinear shifts of the frequencies~\cite{Flach:ArxivRep2014-1}. 
This leads to a nonlinear mechanism of energy exchange between the localized and extended modes in the spectrum (see Section~\ref{numericalspectrum}), which in turn implies that energy that was previously stored in localized modes can now be carried through the system by being transferred either to other localized modes that are spatially close to the original one or to extended modes. In short, there is more transport. Consequently, the force is distributed among a larger number of particles in the chain. This effect is analogous to phenomena that have been observed in disordered NLS
and KG lattices \cite{Flach:ArxivRep2014-1}, and analogous dynamics has also been observed experimentally in the context of waveguide arrays \cite{Heinrich:NJP2012,Naether:OL2013}. Remarkably, the localization goes away completely when the precompression goes to $0$, and instead a pure spreading process occurs. 
In other words, the localization phenomenon, in which nearly all of the energy at vanishing precompression would be partitioned into localized traveling waves (which each have a support on only a 
few site of the chain)~\cite{hinch}, is modified drastically because the presence of disorder.


\subsection{Energy Distribution and Second Moment}\label{m2-energies}

As we stated previously, characterizing whether or not dynamics is localized --- and which particular transport properties can characterize localization in a quantitative way --- is a difficult task~\cite{Kramer:RPP1993}, and it has been examined from many different perspectives by several authors. Such methods include (1) computing a localization length~\cite{LL1,LepriLya}, which gives information on how fast the distributions decay; (2) computing finite-time Lyapunov exponents~\cite{Lyapunov1,Johansson:EPL2010} to study KAM tori and chaotic dynamics; (3) directly estimating scaling properties of the energy distribution~\cite{Lepri:PRE2010,Mulansky:NJP2013}; and (4) calculating moments of distributions that are associated with 
the dynamics~\cite{Dunlap:PRL1990,Datta:PRB1995,Kopidakis:PRL2008,Lepri:PRE2010,Naether:OL2013}. The calculation of moments has been especially popular, and it is particularly common to investigate the growth of the second moment as a function of time, as this gives information about the width of a distribution. However, the exclusive use of the second moment as a single-parameter description is problematic and can lead to a misunderstanding of a system's actual dynamics~\cite{Flach:ArxivRep2014-1,Flach:ArxivRep2014-2}, particularly in strongly nonlinear situations. Consequently, 
following~\cite{Flach:ArxivRep2014-1,Flach:ArxivRep2014-2}, in the present work, we examine dynamics by computing not only the second moment but also the IPR (see Section~\ref{diff-nonlin}). 

Proceeding with our analysis, we note that the total energy of the system is conserved by the dynamics. We are thus interested in the energy distribution's second moment
\begin{equation}
	\tilde{m}_2(t) = \frac{\sum_n (n-n_c)^2 E_n}{\sum_n E_n}\,,
\end{equation}
where $E_n$ is the energy density of the $n$th particle and $n_c$ is the position of the center of the distribution. The energy density of the $n$th particle is given by
\begin{equation}
	E_n(t) = K_n(t)+ V_n(t)\,,
\label{energy}
\end{equation}
where
\begin{equation}
	K_n(t) =\frac{m_n}{2}\dot{u}_n^2(t)
\end{equation}
is the particle's kinetic energy and the potential energy $V_n$ depends on the model. For example, in the linear limit, the potential energy is
\begin{align}
	 V_n(t) &=
	\frac{1}{2}\left[\frac{B_{n}}{2}\left(u_{n-1}(t)-u_{n}(t)\right)^2 \right.\nonumber\\&\qquad\left. +\frac{B_{n+1}}{2}\left(u_{n}(t)-u_{n+1}(t)\right)^2\right]\,.
\end{align}
In the weakly nonlinear regime, the potential energy is
\begin{widetext}
\begin{align}
	V_n^{W}(t) &=
\frac{1}{2}\sum_{i=1}^3
\left[\frac{B_{n}^{(i)}}{(i+1)}(u_{n-1}(t)-u_{n}(t))^{(i+1)}+
	\frac{B_{n+1}^{(i)}}{(i+1)}(u_{n}(t)-u_{n+1}(t))^{(i+1)}\right]\,.
\end{align}
In the strongly nonlinear regime of a Hertzian potential, the potential energy is
\begin{align}\label{hertzpot}
	V_n^{H}(t) &=
\frac{1}{2}
\left\{\frac{2A_{n}}{5}\left[\Delta_{n}+u_{n-1}(t)-u_{n}(t)\right]_{+}^{5/2}+
\frac{2A_{n+1}}{5}\left[\Delta_{n+1}+u_{n}(t)-u_{n+1}(t)\right]_{+}^{5/2}
\right\}\nonumber\\& \qquad-
\frac{1}{2}
\left\{\frac{2A_{n}}{5}\Delta_{n}^{5/2}+
\frac{2A_{n+1}}{5}\Delta_{n+1}^{5/2}
	\right\} - \frac{F_0}{2}\left\{u_{n-1}(t)-u_{n+1}(t)\right\}\,.
\end{align}
\end{widetext}
The two last terms in the right-hand side of the Hertzian potential energy $V_n^{H}(t)$ of Eq.~\eqref{hertzpot} have minus signs, so $V_n^{H}(t) \rightarrow V_n^{W}(t)$ in the weakly nonlinear limit and $V_n^{H}(t) \rightarrow V_n(t)$ in the linear limit. 
The first term in Eq.~\eqref{hertzpot} gives only a trivial contribution to the total energy, because it corresponds to the (constant) background energy associated with the precompression. The last term in Eq.~\eqref{hertzpot} is a telescopic series when one considers all $n$, and the boundaries do not play any significant role because we are interested in the bulk dynamics. The displacement and the momentum at the edges of the chain are both exactly $0$ for all times. 

In the linear regime and in the absence of disorder, the only possible situation after a very long time is for the system to thermalize~\cite{Lepri1,Lepri2}, so one obtains equipartition of energy between the different degrees of freedom.  As a result (and as is well-known), the asymptotic spreading dynamics in a homogeneous chain is ballistic 
(i.e., $\tilde{m}_2(t)\sim t^2$ as $t \rightarrow \infty$) regardless of whether the initial condition is a local displacement perturbation (i.e., $\{u_n(0),\dot{u}_n(0)\} \propto \{\delta_{n,n_c},0\}$) or a local velocity perturbation (i.e., $\{u_n(0),\dot{u}_n(0)\} \propto \{0,\delta_{n,n_c}\}$)
~\cite{Landau:StatPhys,Datta:PRB1995}. 
However, introducing either disorder or nonlinearity can drastically change transport properties \cite{Flach:ArxivRep2014-1}. For example, attempting to estimate a scaling relationship for the second moment now typically produces a different exponent: $\tilde{m}_2(t) \sim t^{\gamma}$ as $t \rightarrow \infty$, where $\gamma \neq 2$.

However, one can expect even more complicated phenomena, so in particular it is not always meaningful to fit the spreading of the second moment to a power law with a single exponent~\cite{Lepri:PRE2010}. 
When there is reasonable power-law scaling, the behavior is called ``superdiffusive'' when $\gamma \in (1,2)$, ``diffusive'' when 
$\gamma = 1$, and ``subdiffusive'' when $\gamma \in (0,1)$.  There is no diffusion when $\gamma = 0$. 
Following the work by Lepri et. al.~\cite{Lepri:PRE2010}, we attempt to identify the situations in which it is reasonable to construe the second moment as having a power-law scaling by 
using as a diagnostic the logarithmic derivative,
\begin{equation}\label{logderivative}
	L_d = \frac{d (\ln \tilde{m}_2(t))}{d (\ln t)}\,,
\end{equation}
where we calculate $\tilde{m}_2$ as a mean over some number of different realizations of the disorder.
We expect that $L_d(t)\rightarrow \gamma$ when $\tilde{m}_2(t)\sim t^{\gamma}$ as $t \rightarrow \infty$, but that $L_d(t)$ can exhibit oscillations when the dynamics are more complicated.
In our numerical computations, we estimate the logarithmic derivative using the finite-difference 
approximation $L_d \approx \Delta (\ln \tilde{m}_2(t))/\Delta (\ln t)$, where we discretize time as in our numerical integration. 
The criterion that we use to state when $\tilde{m}_2$ has a power-law scaling is 
\begin{equation}\label{criterion}
	|L_d(t) - \gamma|<\kappa\,, \quad \mbox{for all} \quad  t>t^*\,, 
\end{equation}
with $\kappa$ a small parameter and $t^*$ an arbitrary time within our observation horizon. We thereby separate the cases in which 
oscillations of the numerical data for the second moment are admissible as statistical fluctuations from the ones in which oscillations are larger than statistical fluctuations. 
 
It is also useful to compute the spectral density associated with the dynamics, as that allows one to identify which frequencies are involved in the dynamics~\cite{Naether:PRE2013}. We use the spatiotemporal displacement distribution to calculate the normalized spectral density 
\begin{equation}
 	g(\nu) = \frac{\sum_n \bar{u}_n^2(\nu)}{\text{max}\left\{\sum_n\bar{u}_n^2(\nu)\right\}}\,,
\label{spectraldensity}
\end{equation}
where 
\begin{equation*}
	\bar{u}_n(\nu)\equiv \sum_{k=0}^{K-1}u_n(t_k)e^{-2\pi i \nu t_k/T_{\mathrm{max}}}\,,
\end{equation*}	
and we use the time points $\{t_k\}_{n=0}^{K-1}$ to partition the interval $[0,T_{\mathrm{max}}]$ into uniform subintervals. 

In Sec.~\ref{numericalspectrum}, we discussed the effects of disorder in strongly precompressed chains of spheres, and we showed that disorder splits the spectrum into a low-frequency region (in which the modes are extended) and a high-frequency region (in which modes tend to be localized). We now seek to explore the interplay between disorder and nonlinearity in both the strongly-precompressed (i.e., weakly nonlinear) regime and the strongly nonlinear regime (whose limiting case is a sonic vacuum). We integrate Eq.~\eqref{Hertz} numerically using a ``$\text{SABA}_{2}\text{C}$'' algorithm \cite{Laskar:Celest2001,Skokos:PRE2009,skokos-erratum}, which is a symplectic integrator that allows one to conserve energy for long temporal evolution. Using $\text{SABA}_{2}\text{C}$, the relative error in energy is between $\Delta E \approx 10^{-9}$ and $\Delta E \approx 10^{-7}$ (depending on the simulation parameters) using a reasonably small time step of $\tau \approx 1$ $\mu$s.

\begin{figure*}[th!]
\includegraphics[width=16.cm]{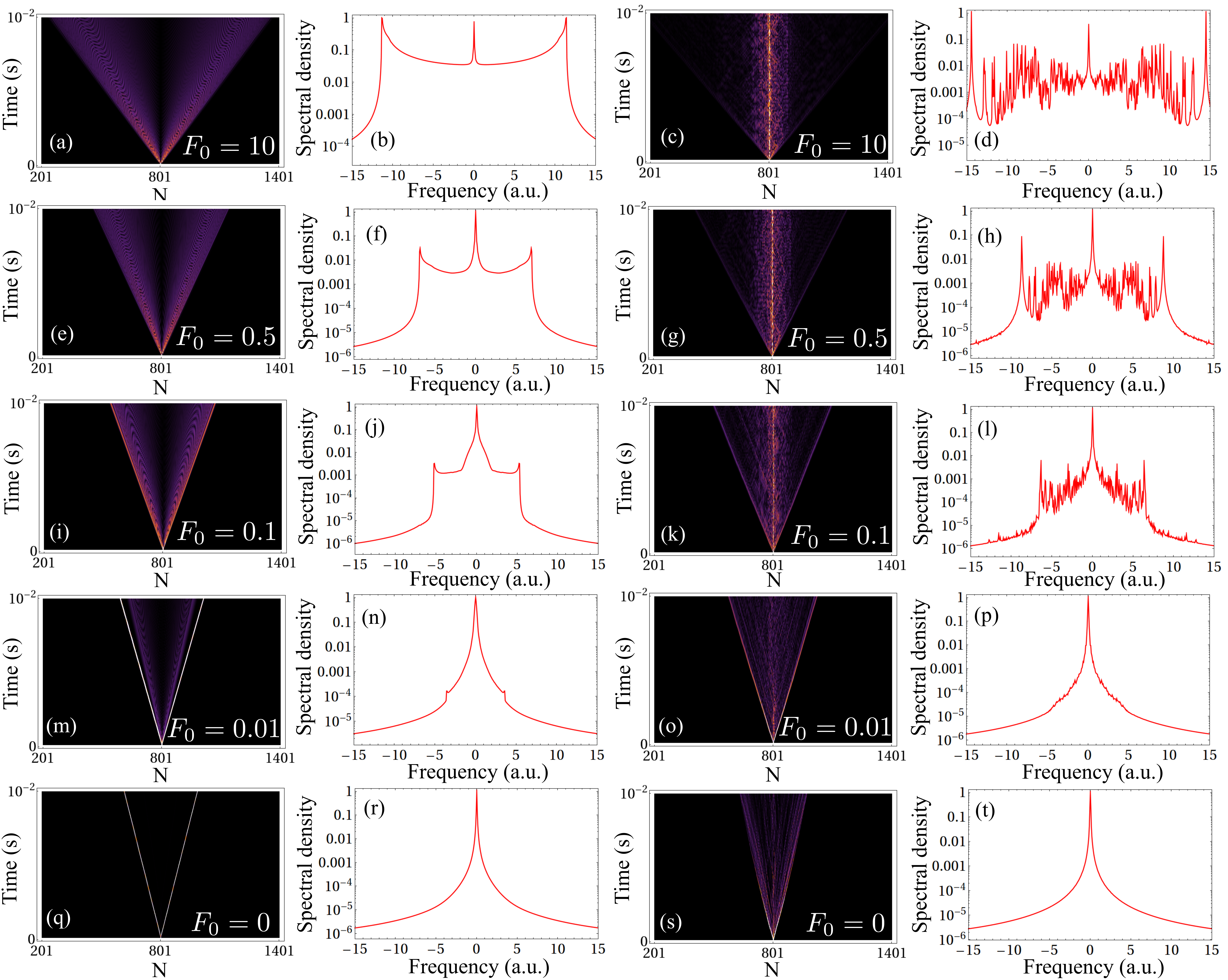}
\caption{(First and third columns) Absolute values of the spatiotemporal energy distributions and (second and fourth columns) spectral density for the dynamics of an initially localized displacement 
perturbation $\{u_n(0),\dot{u}_n(0)\}_{I}=\{\alpha\,\delta_{n,801},0\}$, with $\alpha = 10^{-1}$ $\mu$m, for different amounts of precompression. The first two columns are for a homogeneous chain, and the last two columns are for an 
Anderson chain. Each chain has $N=1601$ particles, though we only show the central $1201$ particles in our plots of spatiotemporal energy distributions. For each example, the integration time is $T_{\mathrm{max}}=10^{-2}$ s, 
and we give the force in units of Newtons. For each row, all panels are for a chain with the same specified precompression strength.}
\label{f3-fourier}
\end{figure*}

\begin{figure*}[th!]
\includegraphics[width=16.cm]{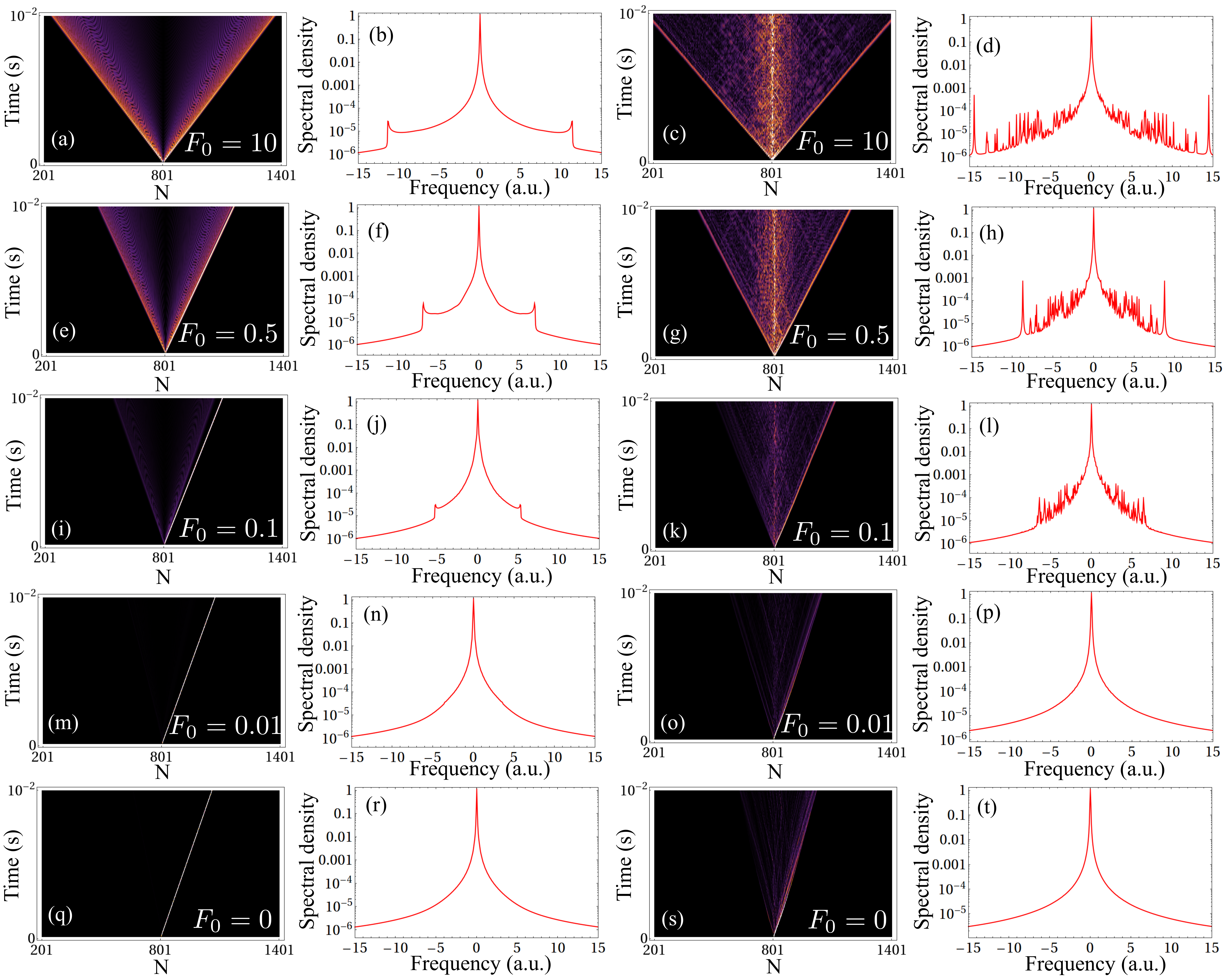}
\caption{(First and third columns) Absolute values of the spatiotemporal energy distributions and (second and fourth columns) spectral density for the dynamics of an initially localized velocity 
perturbation $\{u_n(0),\dot{u}_n(0)\}_{II}=\{0,\beta\,\delta_{n,801}\}$, with $\beta=8\times 10^{-3}$ m/s, for different amounts of precompression. The first two columns are for a homogeneous chain, and the last two columns are for an 
Anderson chain. Each chain has $N=1601$ particles, though we only show the central $1201$ particles in our plots of spatiotemporal energy distributions. For each example, the integration time is $T_{\mathrm{max}}=10^{-2}$ s, 
and we give the force in units of Newtons. For each row, all panels are for a chain with the same specified precompression strength.}
\label{f3-fourierv}
\end{figure*}


\subsubsection{Displacement-Perturbation Initial Conditions}\label{displacement}

In Fig.~\ref{f3-fourier}, we show the spatiotemporal energy distribution and the spectral density for both homogeneous and Anderson-like disordered chains for different levels of precompression and for 
a displacement-perturbation initial condition $\{u_n(0),\dot{u}_n(0)\}_{I}=\{\alpha\,\delta_{n,n_0},0\}$, with  $\alpha = 10^{-1}$ $\mu$m. For $F_0=10$ N, we see that the main contribution to the dynamics comes from the 
linear modes (as we discussed previously). For the homogeneous case, maxima at nonzero frequencies give the band-edge frequencies, where the linear spectrum is denser than it is near $0$ frequency. 
When one decreases the precompression, the band width decreases, and the spreading of waves from the linear modes becomes slower because the sound speed also decreases. One observes clear nonlinear pulses in the dynamics, and 
the speed of these pulses is larger than the sound speed for sufficiently small precompression. (See, for instance, the panels in Fig.~\ref{f2-profiles} with $F_0\leq 0.01$ N.) However, for $F_0\rightarrow 0$, the localized initial condition 
splits into traveling pulses that propagate in opposite directions. This occurs because all of the frequencies of the linear spectrum tend to $0$ for $F_0\rightarrow 0$. 

The chain with Anderson-like disorder exhibits more complicated dynamics than the homogeneous chain. In Fig.~\ref{f3-fourier}, we observe Anderson-like localization for strong levels of precompression.
Spikes in the spectral density indicate the modes that contribute the most to the dynamics. The highest spike is located at a high frequency, so the main contribution comes from a localized mode 
(see Fig.~\ref{f8:linearspectrum}) that is presumably close to the position of the initial excitation. As we can see from the low-frequency spikes in the spectral density, the localization process occurs on top of a diffusive background pattern that arises
primarily because of extended modes.
As we consider weaker precompression, we observe a narrower frequency band near $0$ frequency in the spectral density, analogous to our observations for homogeneous chains. Although there have been many efforts to study the interplay between disorder and 
nonlinearity --- and their effect on spreading dynamics --- most prior research has concentrated on weakly nonlinear settings. 
In fact, the majority of prior work has concentrated on NLS and KG lattices (see, e.g.~Ref.~\cite{Flach:ArxivRep2014-1} and references therein). It has been observed in these settings that 
transport is typically subdiffusive. A notable example in which neighboring lattice sites are not coupled linearly was investigated recently in Ref.~\cite{Mulansky:NJP2013}, who considered strongly nonlinear lattices in which both the on-site and the inter-site interactions are nonlinear. However, those systems also exhibits subdiffusive spreading. 
We believe that the contribution of the on-site nonlinearity is crucial for obtaining subdiffusive spread in lattices with Anderson-like disorder, as the energy-spreading exponents 
in~\cite{Mulansky:NJP2013} differ considerably from the ones that we identify in the present work. Although effects from nonlinearity and disorder can separately localize energy --- and, 
indeed, that is their general predilection, as we can see in Figs.~\ref{f3-fourier}(c,q) --- exactly the opposite can occur in some situations that 
include both of these factors [see Figs.~\ref{f3-fourier}(o,s)]. In particular, we find when both disorder and nonlinearity are present that it is possible for spreading to be enhanced rather than for the two features to conspire to create additional localization. 

For granular chains in the strongly nonlinear regime, neither localization 
in the form of intrinsic localized modes nor exact localization as traveling 
nearly-compact waves is possible, as each of these structures is destroyed by disorder. It is
also impossible to localize in an Anderson-like way, as such localization is suppressed by
nonlinearity and the absence of a linear limit.  Instead, the energy spreads among the particles in a peculiar but characteristic way: strongly localized (and nearly compact) waves are still 
present at the edges of the energy distribution during the spreading process at $F_0=0.001$ N [see Fig.~\ref{f3-fourier}(o)]; however, for $F_0=0$ N, the disorder induces multiple scattering events, which causes the wave amplitudes to decrease [see Fig.~\ref{f3-fourier}(s)].


\subsubsection{Velocity-Perturbation Initial Conditions}\label{velocity}

To analyze the dynamics for an initial velocity perturbation, we consider $\{u_n(0),\dot{u}_n(0)\}_{II}=\{0,\beta\,\delta_{n,n_0}\}$, and we set these perturbations to have the same energy as with the initial displacement perturbation $\{u_n(0),\dot{u}_n(0)\}_{I}=\{\alpha\,\delta_{n,n_0},0\}$.
To get $\beta$ as a function of $\alpha$ (or vice-versa) one needs to solve
$\left.\sum_{n}E_n\right|_{I} = \left.\sum_{n}E_n\right|_{II}$. For example, to express the velocity perturbation in terms of the displacement perturbation, we write
\begin{equation}
	\beta = \sqrt{\frac{4 A}{5m}\left(\left[\Delta-\alpha\right]_+^{5/2}+\left[\Delta+\alpha\right]_+^{5/2}-2\Delta^{5/2}\right)}\,.
\end{equation}
Thus, in our numerical simulations, we set $\beta \approx 8\times 10^{-3}$ m/s, which is the value that we obtain for a homogeneous chain with $F_0 = 10$ N and $\alpha = 10^{-1}$ $\mu$m.

In Fig.~\ref{f3-fourierv}, we show spatiotemporal energy distributions and spectral density for both homogenous and Anderson-like chains using an initially localized velocity perturbation. 
The main --- and fundamental --- difference compared to what we observed using displacement-perturbation initial conditions (see Fig.~\ref{f3-fourier}) comes from the spectral density. When there is 
strongly precompression, we observe that the distribution of modes that are excited by the velocity-perturbation initial condition is denser near $0$ frequency than it is elsewhere. 
In the disordered case, this implies that the mean contribution to the dynamics comes from extended modes rather than localized modes. This contrasts starkly with our observations using 
displacement perturbations, and it leads to dynamics in which the energy spreads much faster than for displacement excitations. Moreover, for velocity perturbations, the energy that diffuses 
in the background is comparable to the amount of energy that remains localized. For $F_0 = 0.1$ N, we observe in both homogeneous and Anderson-like chains that a solitary wave propagates faster 
than the spreading pattern [see Figs.~\ref{f3-fourierv}(i,k)].
For weaker precompression, the solitary wave still propagates in the homogeneous chain, but its amplitude decays in an Anderson-like chain. In particular, when $F_0\rightarrow 0$, the solitary waves are delocalized due to scattering with defects in the disordered chain, and the energy pattern that emerges is qualitatively similar 
to what was observed in \cite{Ponson:PRE2010} for transport of solitary waves in the RDM2 case in a high-disorder regime.

\begin{figure}[th!]
\includegraphics[width=6.5cm]{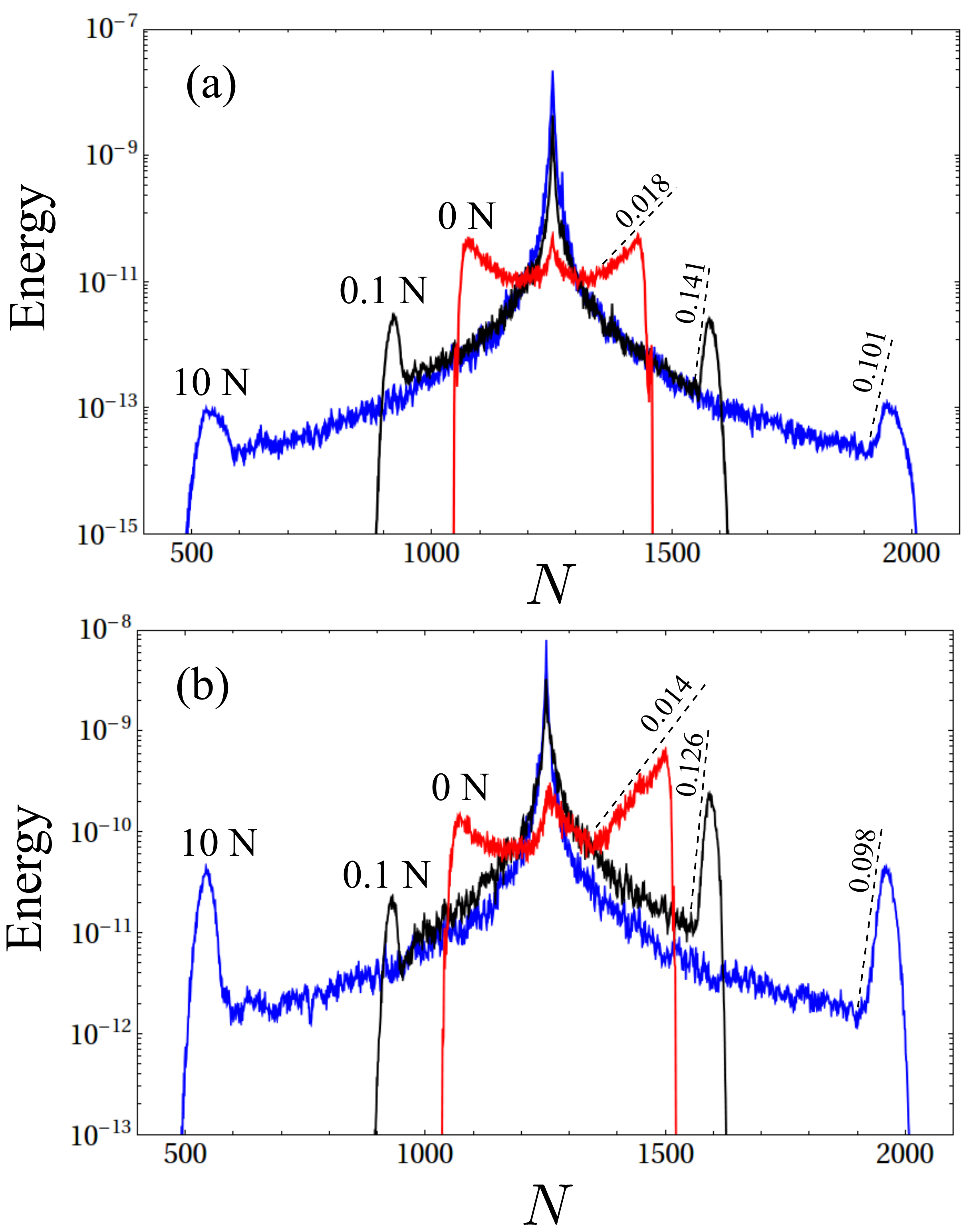}
\caption{Averaged energy distribution for different levels of precompression ($F_0 = 0$, $F_0 = 0.1$, and $F_0 = 10$ N) at $t=10^{-2}$ s for Anderson-like chains. We show the energy using a logarithmic scale, and we average our results over 100 realizations of the disorder for $\xi = 0.5$, $q=0.5$, and $N=2501$. (a) Displacement-perturbation initial condition
($\{u_n(0),\dot{u}_n(0)\}_{I}=\{\alpha\,\delta_{n,1251},0\}$, with $\alpha = 10^{-1}$ $\mu$m); and (b) velocity-perturbation initial condition ($\{u_n(0),\dot{u}_n(0)\}_{I}=\{0,\beta\, \delta_{n,1251}\}$, with $\beta=8\times 10^{-3}$ m/s). Using dashed lines, we show the exponent for the exponential growth of the energy distribution at the chain edges.}
\label{eprofile}
\end{figure}

To visualize what happens to the energy from the dynamics in Anderson-like chains, we average the energy distribution at $t=10^{-2}$ s over 100 realizations. In Fig.~\ref{eprofile}, we show this energy distribution using a logarithmic scale for both displacement excitations and velocity excitations. We examine how the distribution changes depending on the strength of nonlinearity. Specifically, we observe that the energy distribution grows exponentially near the edges of a chain. For $F_0 = 10$ and $F_0 = 0.1$ N, this occurs in a narrow region (fewer than 30 sites) of the chain, and energy is localized at the edge of the distribution 
because traveling waves survive the disorder. This phenomenon is considerably less prominent for displacement-perturbation initial conditions than for velocity-perturbation initial conditions. For $F_0=0$ N (sonic-vacuum regime), we also observe exponential growth of the energy distribution near the chain edges. In this case, however, it occurs over a wider region (about 100 sites for displacement excitations and about 150 sites for velocity excitations), and the exponential growth has a considerably lower exponent than in the chains with nonzero precompression.

\begin{figure}[th!]
\includegraphics[width=8.7cm]{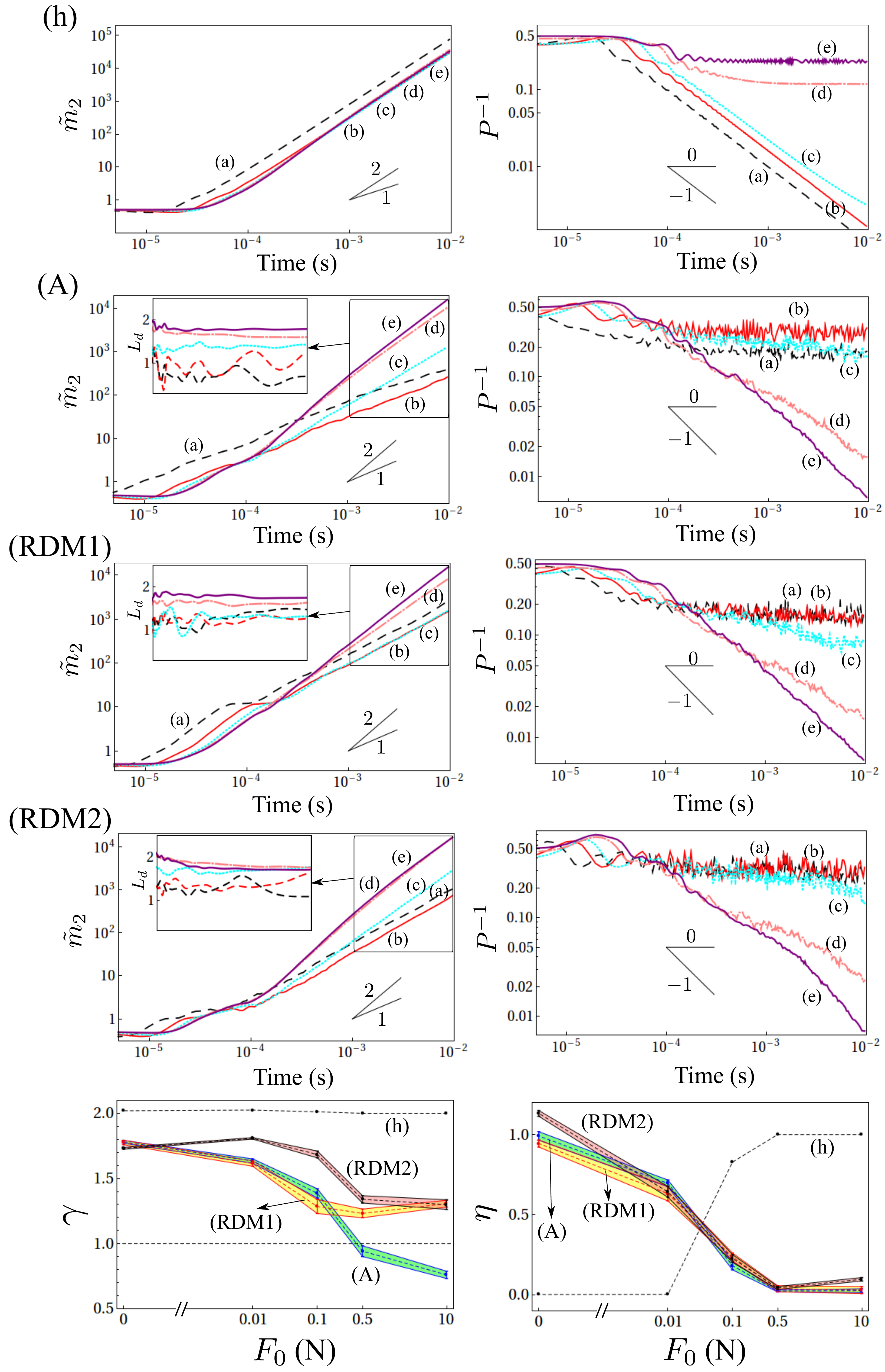}
\caption{Log-log plots of (left) the second moment and (right) IPR as a function of time. Each row is associated with a certain type of disorder (or lack thereof): ``A'' for the Anderson-like chain, ``d'' for RDM1, ``s'' for RDM2, and ``h'' for a homogeneous chain. In each panel, the colors and labels indicate different amounts of precompression $F_0$: (a, dashed black) $10$ N, (b, solid red) $0.5$ N, (c, dotted cyan) $0.1$ N, (d, dash-dotted pink) $0.01$ N, and (e, solid purple) $0$ N. 
To guide the eye, we show slopes of $2$ (ballistic transport) and $1$ (diffusive transport) for the second moment $\tilde{m}_2$ and slopes of $0$ and $-1$ for the IPR $P^{-1}$. In all cases, we use 
chains with $N=2501$ spheres, and the initial condition is $\{u_n(0),\dot{u}_n(0)\}_{I}=\{\alpha\,\delta_{n,1251},0\}$, with $\alpha = 10^{-1}$ $\mu$m.
For the Anderson-like, RDM1, and RDM2 chains, we use the parameter values $\xi=0.5$ and $q=0.5$, and we average over 500 different realizations of a disordered configuration in each case. In the 
last row, we show exponents $\gamma$ and $\eta$ that we obtain for $t\in[4, 10]$ ms by fitting the data using the relations $\tilde{m}_2\sim t^{\gamma}$ and 
$P^{-1}\sim t^{-\eta}$. The insets show the (discretized) logarithmic derivative of the second moment for $t\in[1, 10]$ ms. 
}
\label{f13-m2ipr}
\end{figure}

\begin{figure}[th!]
\includegraphics[width=8.7cm]{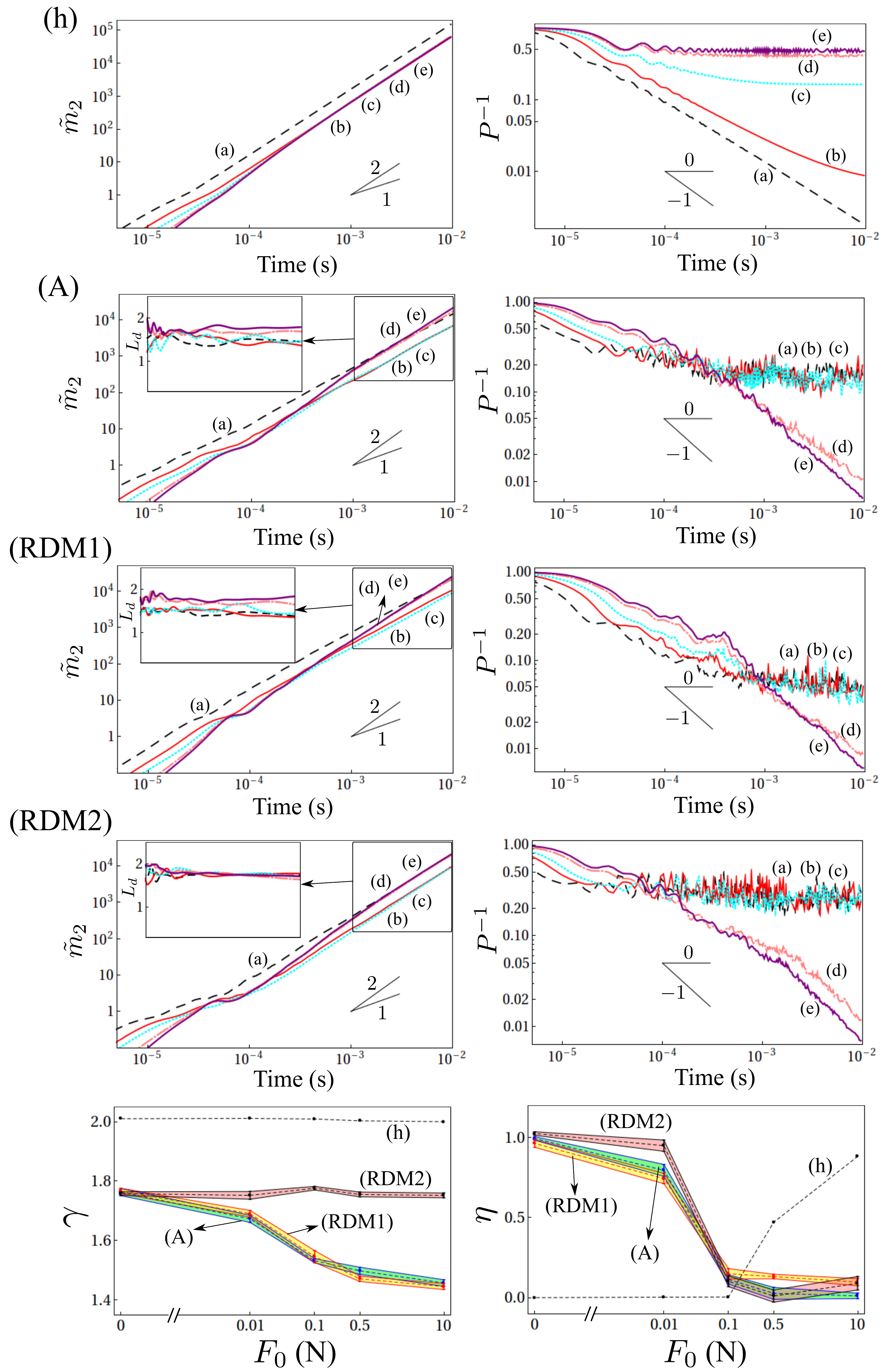}
\caption{Log-log plots of (left) the second moment and (right) the IPR as a function of time. Each row is associated with a certain type of disorder (or lack thereof): ``A'' for the Anderson-like chain, 
``d'' for RDM1, ``s'' for RDM2, and ``h'' for a homogeneous chain. In each panel, the colors and labels indicate different amounts of precompression $F_0$: (a, dashed black) $10$ N, 
(b, solid red) $0.5$ N, (c, dotted cyan) $0.1$ N, (d, dash-dotted pink) $0.01$ N, and (e, solid purple) $0$ N. To guide the eye, we show slopes of $2$ (ballistic transport) and $1$ (diffusive transport) 
for the second moment $\tilde{m}_2$ and slopes of $0$ and $-1$ for the IPR $P^{-1}$. In all cases, we use chains with $N=2501$ spheres, and the initial condition is 
$\{u_n(0),\dot{u}_n(0)\}_{II}=\{0,\beta\,\delta_{n,1251}\}$, with $\beta=8\times 10^{-3}$ m/s. For the Anderson, RDM1, and RDM2 chains, we use the parameter values $\xi=0.5$ and $q=0.5$, and we 
average over 500 different realizations of a disordered configuration in each case. In the last row, we show exponents $\gamma$ and $\eta$ that we obtain for $t\in[4, 10]$ ms by fitting the data using the relations $\tilde{m}_2\sim t^{\gamma}$ and $P^{-1}\sim t^{-\eta}$. The insets show the (discretized) logarithmic derivative of the second moment for $t\in[1, 10]$ ms. 
}
\label{f13-m2iprv}
\end{figure}

\subsection{Transport Arising from Nonlinearity}\label{diff-nonlin}

To quantitatively characterize transport and localization processes, we conduct long-time simulations --- up to $T_{\mathrm{max}}=0.01$ s --- in chains with $N=2501$ spheres. We use long 
chains to avoid boundary effects during the entire numerical integration; no waves reach the boundary of the system within the simulation time. We compute the second moment $\tilde{m}_2$ and 
the IPR $P^{-1}$ as functions of time for the three types of disorder, and we average our results over 500 realizations of a chain configuration in each case. To confirm our numerical results, 
we conduct several tests. For example, we compare our results from $\text{SABA}_{2}\text{C}$ with those using a Runge--Kutta scheme with a very small time step (between $\tau = 0.01$ $\mu$s and 
$\tau = 0.001$ $\mu$s), and we obtain quantitatively the same results for the same realization of disorder \footnote{Importantly, using $\text{SABA}_{2}\text{C}$ allows much longer simulation 
times and a significant improvement in energy conservation in 
comparison to using a Runge--Kutta scheme.}. We also test the $\text{SABA}_{2}\text{C}$ scheme using smaller time steps ($\tau = 0.1$ $\mu$s and $\tau = 0.01$ $\mu$s) and larger system sizes 
($N = 5000$ and $N = 10000$ particles), and we again obtain the same results. In the current section, we compute the IPR [see Eq.~\eqref{IPReq}] using the energy distribution instead of the displacement
distribution. In other words, $h(v_n,\dot{v}_n) = E_n(t)$, and $\tilde{m}_2$ and $P^{-1}$ are also based on the energy distribution.

In Fig.~\ref{f13-m2ipr}, we plot the second moment $\tilde{m}_2$ and the IPR as functions of time for the initial condition with displacement $u_n=10^{-1}\times \delta_{n,1251}$ $\mu$m and all 
particles having speeds of $0$. We also plot $\tilde{m}_2$ and the IPR as functions of time in Fig.~\ref{f13-m2iprv}, but now we use an initial condition with velocity 
$\dot{u}_n=8\times 10^{-1}\times \delta_{n,1251}$ m/s and all particles starting from the equilibrium position. For both cases, we also calculate the (discretized) logarithmic derivative of the second
moment [see Eq.~\eqref{logderivative}] for $t\in[1, 10]$ ms as a diagnostic to test for power-law scaling. 
In most cases, we observe that the scaling $\tilde{m}_2 \sim t^{\gamma}$ persists --- where the exponent $\gamma$ depends strongly on the external force $F_0$ and on the type of disorder --- according to our criterion~\eqref{criterion} and within our observation time. 

However, for strong precompression and displacement-perturbation initial conditions, the second moment involves oscillations that are larger than statistical fluctuations, thereby rendering it impossible to identify a specific power-law trend for the second 
moment in these cases.
These oscillations arise for all three types of disorder, but they are larger for Anderson-like and RMD2 chains than for RDM1 chains. Similar behavior was 
observed by Lepri et al.~\cite{Lepri:PRE2010} for FPU chains with Anderson-like disorder. Their system is similar to our weakly nonlinear regime, but it is not precisely the same: in the FPU chains from \cite{Lepri:PRE2010}, disorder arises only in the linear coupling terms; in our case, disorder arises in nonlinear coupling terms [see, e.g., Eq.~\eqref{HertzWNL}]. 
We estimate $\gamma$ and $\eta$ by taking log-log plots and fitting the numerical data between $4\times 10^{-3}$ s and $10^{-2}$ s 
with a linear function. Specifically, we use the conjugate gradient method and we fit 
for different intervals of time between $4\times 10^{-3}$ s and $10^{-2}$ s, then we average the slopes and calculate the standard deviation, which we estimate as the 
exponents and their error, respectively.
We show our estimates for different values of the precompression 
in Tables~\ref{tab:table1} (for $\gamma$) and~\ref{tab:table2} (for $\eta$) and also graphically in the last row of Figs.~\ref{f13-m2ipr} and~\ref{f13-m2iprv}.
We also attempt to estimate a value of $\gamma$ even for the cases in which the trend of the second moment is more complicated than a power law. We highlight 
these cases using the symbol ``*'' in Table~\ref{tab:table1}, and we stress that the reported exponents correspond to mean values of $L_d(t)$ for $t \in [4, 10]$ ms.

We also compute the second moment and IPR for several other combinations of the parameters $\xi$ and $q$, and we find the same qualitative behavior: the transport is superdiffusive, and weaker precompression yields increased transport. However, we observe that the time required for the system to reach its asymptotic behavior depends on the parameters $\xi$ and $q$ in a nontrivial way, and it is faster for $q\geq 0.5$ in most of the cases that we tested.

A remarkable result is that, in the sonic-vacuum regime, the transport exponents are roughly independent of both the type of disorder and the type of initial condition. We obtain $\gamma\approx 1.7$ and $\eta\approx 1$. It seems that the mechanism that underlies the superdiffusive dynamics in the sonic-vacuum regime may also be independent of the disorder and the initial condition. However, further research in this direction is necessary to truly understand the mechanisms that yield the dynamics in this regime. 

We now summarize the principal results of our numerical simulations on the effect of nonlinearity on energy spreading. For all three types of disorder, the transition from strong precompression 
to weak precompression yields an increase in the diffusivity, as one can see from the increase of the exponent $\gamma$. Perhaps even more importantly, we find that the behavior is typically 
{\it superdiffusive}. For RDM1 and RDM2 chains, we observe superdiffusive transport for all of the precompression strengths that we consider. By contrast, for an Anderson-like chain, we observe 
that the spreading rate depends on the type of initial condition. It is superdiffusive for all precompression strengths for velocity perturbations, whereas we observe superdiffusive transport only for weak precompression for displacement perturbations. 
For strong precompression, our criterion~\eqref{criterion} is
not satisfied. Moreover, for sufficiently strong precompression (see, e.g., Fig.~\ref{f13-m2ipr} for an Anderson-like disorder at $F_0=10$ N), the spreading has slowed down to the point that $L_d(t)<1$. In other words, the spreading has become subdiffusive.

 The dynamics of disordered granular chains depart substantially from the principally subdiffusive behavior that was identified previously in the KG and NLS lattices~\cite{Flach:ArxivRep2014-1,Flach:ArxivRep2014-2} and 
even in the strongly nonlinear lattices of~\cite{Mulansky:NJP2013}. It is likely that the considerably enhanced diffusivity that we observe arises from the FPU nature of our lattices, as FPU and FPU-like lattices are significantly more conducive to traveling waves than, e.g., the DNLS or KG lattices that have constituted the bulk of the settings in previous studies of nonlinear disordered lattices. The observed asymptotic behavior of the IPR also illustrates a form of delocalization in which the energy is no longer split into solitary traveling waves as it is in homogeneous chains. The IPR scaling exponent $\eta$ decreases as $F_0$ increases, which implies in turn that there is an increase in the number of particles that experience large-amplitude vibrations. This is entirely contrary to the expectation for the sonic-vacuum regime in the homogeneous limit, because the energy
no longer is partitioned into strongly nonlinear, strongly localized
waves. Instead, its spatial distribution is reasonably extended,
{\it despite the absence of linear modes}. The delicate 
interplay of disorder and strong nonlinearity seems to be responsible for this intuitively unexpected outcome.

\begin{table*}[th!]
\caption{\label{tab:table1}%
Numerical estimates for the scaling exponent $\gamma$ for a homogeneous chain (h), Anderson chains (A), RDM1 chains (RDM1), and RDM2 chains (RDM2) using the same data as in Fig.~\ref{f13-m2ipr}. 
Recall that $\gamma$ has been computed using the scaling $\tilde{m}_2 \sim t^{\gamma}$ as $t \rightarrow \infty$, and the asterisks (*) highlight the cases in which $\tilde{m}_2$ behaves markedly different from a power law. 
The sets of columns in the middle and right side of the table represent, respectively, the data associated with displacement-perturbation and velocity-perturbation initial conditions.}
\begin{ruledtabular}
\begin{tabular}{l|cccc|cccc}
\textrm{$F_0$ (N)}&\textrm{(h)}&\textrm{(A)}&\textrm{(RDM1)}&\textrm{(RDM2)}&\textrm{(h)}&\textrm{(A)}&\textrm{(RDM1)}&\textrm{(RDM2)}\\
\colrule
$10$ &     $2.000$  &       $0.759 \pm 0.028 (*)$ & $1.306 \pm 0.026$ &  $1.299 \pm 0.037 (*)$&     $2.000$  &       $1.457 \pm 0.011$ & $1.445 \pm 0.010$ &  $1.752 \pm 0.008$\\
 $0.5$ &     $2.000$&        $0.941 \pm 0.041 (*)$&  $1.231 \pm 0.033$&  $1.339 \pm 0.028 (*)$&     $2.003$&        $1.498 \pm 0.012$&  $1.471 \pm 0.009$&  $1.755 \pm 0.008$\\
 $0.1$ &     $2.012$&        $1.385 \pm 0.035$&  $1.284 \pm 0.054$&  $1.684 \pm 0.026$&     $2.009$&        $1.532 \pm 0.007$&  $1.546 \pm 0.020$&  $1.776 \pm 0.006$\\
 $0.01$ &    $2.025$&        $1.634 \pm 0.016$&  $1.614 \pm 0.021$&  $1.808 \pm 0.006$&    $2.011$&        $1.673 \pm 0.012$&  $1.690 \pm 0.011$&  $1.752 \pm 0.013$\\
 $0$    &   $2.020$ &       $1.781 \pm 0.011$ & $1.776 \pm 0.017$ & $1.731 \pm 0.007$&   $2.010$ &       $1.763 \pm 0.012$ & $1.767 \pm 0.009$ & $1.758 \pm 0.006$\\
\end{tabular}
\end{ruledtabular}
\label{tabla}
\end{table*}

\begin{table*}[th!]
\caption{\label{tab:table2}%
Numerical estimates for the scaling exponent $\eta$ for a homogeneous chain (h), Anderson chains (A), RDM1 chains (RDM1), and RDM2 chains (RDM2) using the same data as in Fig.~\ref{f13-m2ipr}. 
Recall that $\eta$ has been computed using the scaling $P^{-1}\sim t^{-\eta}$ as $t \rightarrow \infty$. Sets of columns in the middle and right side of the table represent, respectively, the data 
associated with displacement-perturbation and velocity-perturbation initial conditions.}
\begin{ruledtabular}
\begin{tabular}{l|cccc|cccc}
\textrm{$F_0$ (N)}&\textrm{(h)}&\textrm{(A)}&\textrm{(RDM1)}&\textrm{(RDM2)}&\textrm{(h)}&\textrm{(A)}&\textrm{(RDM1)}&\textrm{(RDM2)}\\
\colrule
$10$ &     $1.000$  &       $0.025 \pm 0.012$ & $0.029 \pm 0.022$ &  $0.096 \pm 0.011$&     $0.882$  &       $0.011 \pm 0.017$ & $0.097 \pm 0.022$ &  $0.090 \pm 0.042$\\
 $0.5$ &     $1.000$&        $0.023 \pm 0.006$&  $0.037 \pm 0.019$&  $0.042 \pm 0.009$&     $0.470$&        $0.027 \pm 0.037$&  $0.132 \pm 0.014$&  $0.010 \pm 0.038$\\
 $0.1$ &     $0.828$&        $0.181 \pm 0.026$&  $0.235 \pm 0.025$&  $0.226 \pm 0.024$&     $0.004$&        $0.114 \pm 0.029$&  $0.149 \pm 0.032$&  $0.102 \pm 0.033$\\
 $0.01$ &    $0.002$&        $0.695 \pm 0.021$&  $0.631 \pm 0.046$&  $0.642 \pm 0.034$&    $0.003$&        $0.795 \pm 0.037$&  $0.744 \pm 0.032$&  $0.949 \pm 0.035$\\
 $0$    &   $0.001$ &       $0.991 \pm 0.027$ & $0.944 \pm 0.022$ & $1.132 \pm 0.018$&   $0.000$ &       $0.994 \pm 0.011$ & $0.963 \pm 0.024$ & $1.025 \pm 0.011$\\
\end{tabular}
\end{ruledtabular}
\label{tabla}
\end{table*}

To provide reference values, Tables~\ref{tab:table1} and~\ref{tab:table2} also include our results for the computation of characteristic exponents for homogeneous granular chains.  For linear 
homogeneous chains, one can derive analytically that $\gamma = 2$ and $\eta = -1$ \cite{Datta:PRB1995}. The transport is ballistic for all values of $F_0$, but the IPR saturates below certain 
values of $F_0$. (As one can see in Figs. \ref{f13-m2ipr} and~\ref{f13-m2iprv}, this saturation is clear for $F_0=0.01$ N.) This is related to the system becoming strongly nonlinear, with 
no linear waves propagating, so the initial wave splits into two traveling energy-carrying pulses. These pulses are spatially localized, so $P^{-1}$ does not grow as a function of time.


\section{Conclusions and Discussion}\label{Sec6}

We characterized the localization and transport properties of one-dimensional disordered granular crystals for both 
uncorrelated and correlated disorders. We found, in the linear regime, 
that there are different extended modes that can contribute to the 
transport in a disordered system. We investigated the correlation 
properties of three types of disorder --- an Anderson model and two random 
dimer models --- and we demonstrated that the rules that generate the spin-based dimer 
chain (i.e., RDM2) can contain either short-range or long-range correlations in the disorder.

We showed, by direct diagonalization of the linearized granular chain in the 
presence of precompression, that localized linear modes are mostly 
impurity-like modes and that the spectrum of the linearized chain 
includes a mixture of extended and localized modes. The extended modes 
usually occur at low frequencies. Using a spectral perspective, we again found that RDM2 chains are rather special, as (in contrast to RDM1, which is the traditional ``random dimer model'') they much more closely mirror 
the structure of a perfectly ordered diatomic chain than of a homogeneous chain.

Armed with an understanding of the linear modes, we set out to quantify the nonlinear dynamics of the three different types of disordered lattices. Although the effects of nonlinearity 
(in the absence of disorder) in strongly nonlinear 
homogeneous granular crystals are known to be rather unusual --- 
the energy tends to split into strongly localized traveling pulses, which leads to a saturation of the inverse participation ratio --- we found very surprising and previously
unexplored behavior when one introduces disorder into granular chains. When there is strong precompression (i.e., a very low
level of nonlinearity), disorder tends to localize energy due to Anderson-like effects. Surprisingly, however, 
localization no longer emerges for sufficiently small precompression, 
as a disordered chain tends to a sort of ``thermalization'' as the 
energy spreads throughout the whole chain. In this case, neither the second moment nor the inverse participation ratio saturates, and presumably traveling waves cannot survive the presence of 
disorder in this regime. Nevertheless, it is conceivable in such a setting that stable localized waves may exist due to disorder. 
Furthermore, very recently, \cite{Mobilitydisorder} reported that solitary-wave mobility can be enhanced in certain classes of nonlinear disordered lattices either by specific realizations of a type of disorder or 
with specific initial conditions. It is not clear when the joint presence of disorder and nonlinearity destroys localization, and investigating when this occurs is an important open question. In the sonic-vacuum regime of no precompression, the exponents of the temporal asymptotic scaling of the inverse participation ratio are close to $-1$, which is what occurs in the linear homogeneous case (in particular, for about $F_0 = 10$ N); in other words, the energy is delocalized. However, the transport remains superdiffusive rather than ballistic. In fact, we found that each of our three disordered chain models is {\it typically}
superdiffusive, in stark contrast to what is known about disorder in other lattice 
models~\cite{Flach:ArxivRep2014-1}, in which a self-trapping mechanism always 
dominates as the strength of the nonlinearity increases (independently of the disorder). 

By computing a (discretized) logarithmic derivative $L_d$ of the second moment, we find for strongly precompressed (i.e., weakly nonlinear) chains with initially localized displacement excitations that the spreading does not show a clear diffusive trend. In other words, the second moment behaves in a complicated way and exhibits oscillations that are larger than statistical fluctuations (analogous to what was observed in
\cite{Lepri:PRE2010}). For Anderson-like disorder and $F_0 = 10$ N, we observed that $L_d(t)<1$ during the time interval that we consider. This behavior is clearer in Anderson-like and RDM2 chains
than in RDM1 chains. By contrast, for initially localized velocity excitations, we found that $L_d$ satisfies~\eqref{criterion}, and we thus observed a standard power-law growth for the second moment: $\tilde{m}_2(t)\sim t^{\gamma}$, with $\gamma >1$ (i.e., superdiffusive spreading) for all types of disorder. However, for weakly precompressed (i.e., strongly nonlinear) chains, the dynamics is superdiffusive for all types of disorder and both types of initial
conditions. Surprisingly, in the sonic-vacuum regime, the exponents are very similar (roughly $\gamma\approx 1.7$ and $\eta\approx 1$) and they seem to be independent of both the type of disorder and of whether we use a displacement or velocity excitation as an initial condition. Moreover, granular lattices --- which are inter-site interaction lattices of FPU type --- appear to be far more conducive to energy transport than the previously explored KG and NLS lattices. Presumably, this situation arises from the ability of the granular lattices to transport energy in the strongly nonlinear regime in the form of robust traveling waves.

Our work opens a panorama of both theoretical and experimental possibilities.  From a theoretical perspective, future challenges involve deriving the mechanisms that relate the type of correlation in chain 
disorder to the spectral and transport properties of the system, incorporating 
dissipation (and possibly also restitution) effects into disordered granular crystals, studying higher-dimensional granular crystals, examining other types of initial excitations, and more. 
In particular, we tried initializing the chain at a particular localized mode of the underlying linear system, and we observed that the spreading can be considerably slower in comparison 
to the initial conditions that we used in the present article.   
 However, further investigation is necessary to quantitatively characterize the dynamics. We note, 
however, that the absence of the experimental capability to initialize a granular chain with a specific initial distribution in present settings renders such a study theoretically-motivated rather than practically-motivated 
at the moment.

Indirect experimental measures of the dynamical properties are possible using experimental techniques such as the ones in Ref.~\cite{Boechler:PRL2010}, and (importantly) recent more advanced techniques, such as laser Doppler vibrometry, which now make it possible to measure the spatiotemporal properties of an entire granular chain (see, e.g.,~\cite{cho2}). In such an experimental setup, one can track the force at {\it each} particle as a 
function of time, and one can consequently directly measure quantities such as $\tilde{m}_2$ or
$P$ as a function of time.  Naturally, it is rather tedious to conduct 
experiments with very long chains and using many realizations of a given 
type of disorder. Nevertheless, neither of these is presently out
of reach, as experiments with as many as $188$ particles have been
reported~\cite{Carretero:PRL2009} and multiple realizations to obtain
ensemble averages have also been performed previously (e.g., in~\cite{Ponson:PRE2010}). 
Extending these considerations of the present paper and enabling even a partial measurement
of the exponents presented herein would be extremely exciting in corroborating
the superdiffusive transport dynamics that we have observed for disordered strongly 
nonlinear lattices.


\section*{Acknowledgements}

We thank C. Chong, S. Flach, A. Goriely, I. Hewitt, A. Pikovsky, Ch. Skokos, G. Theocharis, J. Yang, and two anonymous referees for helpful comments. A.~J.~M. acknowledges partial support from CONICYT (BCH72130485/2013).
P.~G.~K. gratefully acknowledges the support of 
NSF-CMMI-1000337, as well as from
the US-AFOSR under grant FA950-12-1-0332  
and the ERC under FP7, Marie
Curie Actions, People, International Research Staff
Exchange Scheme (IRSES-606096).
P.~G.~K.'s work at Los Alamos is supported in part by the U.~S. Department
of Energy.




\end{document}